\documentclass[twocolumn]{aastex62}

\usepackage{graphics,epsf}
\usepackage{amsmath}                
\usepackage{amsfonts}               
\usepackage{amssymb}                
\usepackage{epsfig}   
\usepackage{subfigure}
\usepackage{graphicx}
\usepackage{float}
\usepackage{color}
\usepackage[para,online,flushleft]{threeparttable}

\def \cm{~\rm{cm}}
\def \s{~\rm{s}}
\def \km{~\rm{km}}

\def \K{~\rm{K}}
\def \g{~\rm{g}}

\def \AU{~\rm{AU}}
\def \erg{~\rm{erg}}

\def \yr{~\rm{yr}}
\def \h{~\rm{h}}

\begin{document}

\title{Simulating jets from a neutron star companion hours after a core collapse supernova}


\author{Muhammad Akashi}
\affiliation{Department of Physics, Technion, Haifa, 3200003, Israel; akashi@physics.technion.ac.il; soker@physics.technion.ac.il}
\affiliation{Kinneret College on the Sea of Galilee, Samakh 15132, Israel}

\author[0000-0003-0375-8987]{Noam Soker}
\affiliation{Department of Physics, Technion, Haifa, 3200003, Israel; akashi@physics.technion.ac.il; soker@physics.technion.ac.il}
\affiliation{Guangdong Technion Israel Institute of Technology, Shantou 515069, Guangdong Province, China}


\begin{abstract}
We conduct three-dimensional hydrodynamical simulations to explore the interaction of jets that a neutron star (NS) companion to a type Ic or type Ib core collapse supernova (CCSN) launches few hours after explosion with the ejecta of the CCSN. We assume that a NS companion at $5 R_\odot$ from the exploding star accretes mass from the slower inner ejecta through an accretion disk, and that the disk launches two opposite jets. Although the energy of the jets is only about one per cent of the total energy of the ejecta, it is comparable to the energy of the slower inner part of the ejecta. We find that the jets inflate one low-density hot bubble to one side of the ejecta, and that this bubble expands to influence ejecta gas up to expansion velocities of $v_{\rm ej} \simeq 3500 \km \s^{-1}$. The post-shock jets' material develops a large meridional flow and small vortexes. The boundary between the ejecta and jets' post-shock gases is unstable. The instabilities and vortexes mix jets' gas with the ejecta.
We expect such a hot bubble to contribute to the light curve of CCSNe that are progenitors of binary NS systems, in particular to observers on the side of the bubble.   
\end{abstract}



\section{Introduction}
\label{sec:intro}

Several evolutionary routes of binary massive stars lead to the spiralling-in of a neutron star (NS) inside the envelope of a core collapse supernova (CCSN) progenitor. The discussion to follow holds in most parts also to a black hole spiralling-in inside a massive stellar envelope, but in the present study we simulate non-relativistic jets as expected for NS companions. 
Earlier studies of the common envelope evolution (CEE) of a NS and the CCSN progenitor include the recycling of a pulsar as during the CEE the NS accretes mass with sufficiently high angular momentum (e.g., \citealt{Chattopadhyayetal2020}), and the formation of a close binary system of two NSs that might later emit gravitational waves and merge (e.g., \citealt{Taurisetal2017, Kruckowetal2018, MandelFarmer2018, VignaGomezetal2020}). 

{{{{ The scenario that brings the system to the stage where we begin our simulations is as follows. The initial system is of a binary system of two massive main sequence stars, each star with an initial mass of $\ga 10 M_\odot$, and an orbital separation of $\approx 1 - {\rm few} \times \AU$.
The more massive star evolves first, explodes as a CCSN, and leaves a NS remnant (in principle it can also leave a black hole remnant).
If at explosion the mass of the hydrogen-rich envelope of the primary star is $\ga 1 M_\odot$ it explodes as a type II SN. In some systems, depending on the masses of the stars and the initial orbit, the primary star might lose mass to the secondary star via Roche lobe overflow and lose most of its hydrogen-rich envelope. In that case it will explode as a type IIb SN. If it loses all of its hydrogen rich envelope it will explode as a type Ib SN. Since the system does not enter a CEE at this stage, it is not likely that the primary star will lose its helium, and therefore a type Ic SN at this stage is not likely. 
For the present study it is of no significance which type of CCSN the primary star experiences. 
Later, the initially less massive star, the secondary star, evolves off the main sequence and swallows the NS remnant of the primary star. The NS spirals inside the envelope and removes all the hydrogen rich envelope. The process where the NS accretes mass from the envelope and launches jets might help in removing the envelope (see below). 
}}}}

In most cases where the NS remnant of the primary star does not enter the core of the secondary star the CEE ends with a close binary system of a NS (the remnant of the primary star) and the core of the secondary star, which is a progenitor of a type Ib or a type Ic CCSN, i.e., a stripped-envelope CCSN progenitor (SNe Ibc; e.g., \citealt{Dewietal2002, VignaGomezetal2018, Laplaceetal2020}). 
The typical post-CEE semi-major axis of the NS and the CCSN progenitor might be $a \approx 1-5 R_\odot$, corresponding to an orbital period of $\approx 1-10{~\rm hours}$ (e.g., \citealt{Fragosetal2019, RomeroShawetal2020}). 

{{{{ We comment here on the possibility that the NS launches jets while inside the envelope of the secondary star as it accretes mass from the envelope. }}}}
Mass accretion rates of $\dot M_{\rm acc} \ga 10^{-3} \rm M_\odot \yr^{-1}$ allow the gas that the NS accretes to cool by neutrino emission, therefore facilitating the high mass accretion rates \citep{HouckChevalier1991, Chevalier1993, Chevalier2012}. 
During the CEE the gas that a NS accretes possesses high enough specific angular momentum to form an accretion disk (e.g., \citealt{ArmitageLivio2000, Papishetal2015, SokerGilkis2018, LopezCamaraetal2020}). This accretion disk launches jets. The launching of jets by the mass-accreting NS companion inside the envelope can power a bright transient event, with characteristics similar to those of CCSNe. If the NS does not destroy the core the event is a common envelope jets supernova (CEJSN) impostor (e.g., \citealt{Gilkisetal2019a}) if the NS does destroy the core it a CEJSN (e.g., \citealt{Sokeretal2019, Schroderetal2020, GrichenerSoker2019}).  

{{{{ We return to the binary system of the NS remnant of the primary star and the core of the secondary star, as the later explodes as SNIb or SNIc. }}}} 
When the core of the secondary star explodes and ejecta collide with the NS (e.g., \citealt{EgorovPostnov2009}), the NS companion accretes mass from the CCSN ejecta (e.g., \citealt{Fryeretal2014, Becerraetal2015, Becerraetal2016, Becerraetal2019, Soker2020NSjets}). Because of the orbital motion of the NS relative to the exploding star the accretion flow does not have an axisymmetrical geometry and the accreted mass possesses angular momentum. This is similar to the accretion flow of a NS from a stellar wind. 
There are many studies of NS accreting from a stellar wind (but not from CCSN ejecta) that consider the non-axisymmetrical accretion flow and accretion disk formation (e.g.,  \citealt{IllarionovSunyaev1975, ShapiroLightman1976, Wang1981, DelgadoMarti2001, Erkutetal2019, ElMellahCasse2017, ElMellahetal2018, XuStone2019, Liaoetal2020}).   

In the case of a NS accreting from a CCSN ejecta the gas velocity is not constant. The ejecta velocity profile reaches a homologous expansion, such that the ejecta velocity relative to the NS decreases with time. As a result of that the orbital motion becomes more influential in deviating the flow from axisymmetry, and therefore the specific angular momentum of the accreted gas increases {{{{ (equation 10 in \citealt{Soker2020NSjets}).    }}}}
When the ejecta velocity becomes slow enough, $v_{\rm ej} < v_{\rm ej,d} \simeq 1500 - 1000 \km \s^{-1}$ for $a = 1-5 R_\odot$, respectively \citep{Soker2020NSjets},  the specific angular momentum becomes large enough for the gas to form an accretion disk around the NS (\citealt{Fryeretal2014, Becerraetal2015, Becerraetal2016, Becerraetal2019,Soker2020NSjets}).

In this study we do not simulate the accretion process, but rather assume that by accreting ejecta mass through an accretion disk the NS companion launches jets. We simulate the interaction of these jets with the ejecta. Unlike the case of minutes to hours delayed jets from the newly born NS or black hole that accrete fallback material (e.g.,  \citealt{Stockingeretal2020}), here the flow is not symmetric about the center of the ejecta, and we must conduct three-dimensional (3D) hydrodynamical simulations. We describe the ejecta and jets in section \ref{sec:EjectaJets} and the numerical setting in section \ref{sec:Numerical}. We present our results in section \ref{sec:Results} and summarise in section \ref{sec:Summary}. 

\section{The ejecta and jets}
\label{sec:EjectaJets}

\subsection{The basic flow structure}
\label{subsec:flow}

We take homologous expanding ejecta with a velocity at each radius of $v_{\rm ej}(r)=r/t$ and density of  (\citealt{SuzukiMaeda2019}; their equation 1-6, with $l=1$ and $m=10$), 
\begin{equation}
\rho (r, t) = \begin{cases}
        \rho_0 (t) \left( \frac{r}{t v_{\rm br}} \right)^{-1} 
        & r\leq t v_{\rm br}
        \\
        \rho_0 (t) \left( \frac{r}{t v_{\rm br}} \right)^{-10} 
        & r>t v_{\rm br}, 
        \end{cases}
\label{eq:density_profile}
\end{equation}
    \newline
where $M_{\rm ej}$ is the ejecta mass, $E_{\rm SN}$ is its kinetic energy,  
\begin{eqnarray}
\begin{aligned} 
& v_{\rm br} = \left( \frac{20}{7} \right)^{1/2} \left( \frac {E_{\rm SN}}{M_{\rm ej}} \right)^{1/2} 
=6400 
\\& 
\times \left( \frac {E_{\rm SN}}{10^{51} \erg} \right)^{1/2}
\left( \frac {M_{\rm ej}}{3.5 M_\odot} \right)^{-1/2}
\km \s^{-1} ,
\end{aligned}
\label{eq:vbr}
\end{eqnarray}
and
\begin{eqnarray}
\begin{aligned} 
\rho_0 (t) & = \frac {7 M_{\rm ej}}{18 \pi v^3_{\rm br} t^3} = 7 \times 10^{-5} 
\left( \frac {M_{\rm ej}}{3.5 M_\odot} \right)
\\& \times 
\left( \frac{v_{\rm br}}{6400 \km \s^{-1}} \right)^{-3}
\left( \frac{t}{1 {\rm h}} \right)^{-3}
\g \cm^{-3}.
\end{aligned}
\label{eq:rho0}
\end{eqnarray}
We base our scaling of the ejecta mass of SNe Ibc on \cite{Dessartetal2016} and \cite{Teffsetal2020}, and  on the ejecta mass of SN~Ic~iPTF15dt  \citep{Taddiaetal2019}.
From equation (\ref{eq:density_profile}) the density at radius $r<v_{\rm br} t$  is  
\begin{eqnarray}
\begin{aligned} 
\rho (r,t) & = 4.7 \times 10^{-4} 
\left( \frac{r}{5R_\odot} \right)^{-1} 
\left( \frac {M_{\rm ej}}{3.5 M_\odot} \right)
\\& \times
\left( \frac{v_{\rm br}}{6400 \km \s^{-1}} \right)^{-2} 
\left( \frac{t}{1 {\rm h}} \right)^{-2}
\g \cm^{-3}.
\end{aligned}
\label{eq:rho(rt)}
\end{eqnarray}

We consider a NS of mass $M_{\rm NS}$ and radius $R_{\rm NS}$ that orbits the progenitor of a SN Ibc in a circular orbit of radius $a \simeq 5 R_\odot$ (see section \ref{subsec:Uncertainties} for the evolution with time).
For the parameters we use here the orbital velocity of the two stars with respect to each other is $v_{\rm orb} \simeq 500 \km \s^{-1}$. 
\cite{Soker2020NSjets} finds that the NS companion starts to accrete mass through an accretion disk when the ejecta velocity decreases to about $v_{\rm ej,d} \approx 900 \km \s^{-1}$. 
The relative velocity between the ejecta and companion is 
\begin{equation}
 v_{\rm rel} \simeq [v^2_{\rm ej}(r) + v^2_{\rm orb}]^{1/2}. 
 \label{eq:Vrel}
\end{equation}
When disk formation starts $v_{\rm rel} \simeq 1000 \km \s^{-1}$. 
The Hoyle-Lyttelton accretion radius is 
 \begin{equation}
R_a = \frac{2 G M_{\rm NS}} {v^2_{\rm rel}} =  0.53  
\left( \frac {M_{\rm NS}}{1.4 M_\odot} \right)
\left( \frac {v_{\rm rel}}{1000 \km \s^{-1}} \right)^{-2}  R_\odot. 
\label{eq:Ra}
\end{equation}
\newline
The mass accretion rate is 
\begin{eqnarray}
\begin{aligned} 
& \dot M_{\rm NS}  \simeq \pi R^2_a v_{\rm rel} \rho \simeq 3.7 
\times 10^{-4} \left( \frac{r}{5R_\odot} \right)^{-1} 
\\& \times
\left( \frac {M_{\rm NS}}{1.4 M_\odot} \right)^2
\left( \frac {v_{\rm rel}}{1000 \km \s^{-1}} \right)^{-3} \left( \frac {M_{\rm ej}}{3.5 M_\odot} \right)
\\& \times
\left( \frac{v_{\rm br}}{6400 \km \s^{-1}} \right)^{-2} 
\left( \frac{t}{1 {\rm h}} \right)^{-2}
M_\odot \h^{-1}
\end{aligned}
\label{eq:dotMNS}
\end{eqnarray}
The accretion rate at $t=1 \h$ is therefore $\dot M_{\rm NS,1h}  \simeq 3.7 \times 10^{-4} M_\odot \h^{-1} = 3.2 M_\odot \yr^{-1}$. 

The accretion through a disk starts when ejecta  with velocity of $v_{\rm ej,d} \approx 900 \km \s^{-1}$ reaches the NS,  $t_{\rm acc,0} \simeq 1.1 \h$. After another hour, ejecta  with a velocity of $\simeq 450 \km \s^{-1}$ reaches the NS, after 3 hours it is $\simeq 300 \km \s^{-1}$, and so on. At $t=2 \h$ and $t=3h$ we find that $v_{\rm rel,2} \simeq 670 \km \s^{-1}$ and $v_{\rm rel,3} \simeq 580 \km \s^{-1}$, respectively, and from equation (\ref{eq:Ra}) the accretion radius is larger. By those times the density at the location of the NS are $0.25$ and $0.11$ times the density at $t=1 \h$. The accretion rates at these times are $\dot M_{\rm NS,2h} \simeq 0.83 \dot M_{\rm NS,1h}$ and $\dot M_{\rm NS,3h} \simeq 0.56 \dot M_{\rm NS,1h}$, respectively.

We assume that the jets carry about $3 \% - 10 \%$ of the mass that the NS accretes at the escape velocity from the NS, i.e., $\dot M_{\rm 2j} \simeq 1-3 \times 10^{-5} M_\odot \h^{-1}$ during the 2 hours in the range $t=1-3 \h$. This implies an average power of $\dot E_{\rm 2j} \simeq 3\times 10^{48} - 9 \times 10^{48}\erg \h^{-1}$. To save numerical resources we launch jets at lower velocities, so the mass outflow rate is larger. 

\subsection{Uncertainties and numerical simplifications}
\label{subsec:Uncertainties}

The main processes and their uncertainties are as follows (\citealt{Soker2020NSjets}). 

(1) \textit{Jets' power}. The largest uncertainty is in the {{{{ mass outflow rate into the jets, and therefore also in the }}}} power of the jets. They depend on the mass accretion rate by the NS and the fraction of that mass that the NS launches as jets. 
Another uncertain parameter is the collimation degree of the jets. For example, it is not clear whether the accretion disk is a thin or a thick accretion disk. This might change the collimation degree of the  jets. 
In the present study we concentrate on these uncertainties and vary the jets' power and collimation degree. 
    
(2) \textit{The phase of jets activity.} There is an uncertainties as to the time when the gas that the NS accretes has sufficiently large specific angular momentum to form an accretion disk that launches jets. We do not vary this time in our simulations, and use the estimate from \cite{Soker2020NSjets} that for a pre-explosion orbital radius of $a=5R_\odot$ the formation of an accretion disk starts at about $t \simeq 1 \h$.  
   
(3) \textit{The explosion geometry.} We expect the close NS to spin-up the SN progenitor. Because of that, it is likely that polar jets drive the CCSN explosion. Compared to a spherical explosion, an exploding rapidly rotating core would eject a relatively slower equatorial flow and a faster polar flow  \citep{Gilkisetal2016Super}. A slower equatorial flow results in an earlier formation of an accretion disk and in a higher accretion rate \citep{Soker2020NSjets}.
   
(4) \textit{Post-explosion orbit.} If the system stays bound after explosion the NS acquires a post-explosion eccentric orbit. If the NS escape it acquires a hyperbolic orbit. In any case, after explosion the distance between the NS and the center of explosion increases. The orbit itself is not a trivial one to calculate due to the interaction of the NS with the ejecta. \cite{Becerraetal2015}, for example, found that due to the NS-ejecta interaction the system can stay bound even if the ejecta carry more than $50 \%$ of the initial binary mass.
In the present study we keep the distance between the  NS and the center of explosion constant in time, $a= 5R_\odot$. The NS is moving away from the center of the explosion, but we do not consider it to be the major effect during the two hours of the jets' activity time period. Since we include neither the gravity of the NS nor that of the exploding star, we do not include the orbital motion of the NS. In this study we do not continue to launch jets beyond $t=3 \h$, as the NS moves further away (which we do not include), and therefore the accretion rate substantially decreases, and so is the jets' power.  
  
In addition to these uncertainties we are limited by computer resources. To save computer resources we ignore some effects that introduce errors that are much below the other uncertainties. We reduce the dynamical range (range of gas velocities here) and launch the jets at a velocity of only $v_{j,n}=5 \times 10^4 \km \s^{-1}$ rather than at $\approx 1.5 \times 10^5 \km \s^{-1}$. To have the same jets' power we take the numerical mass outflow rate in the jets, $\dot M_{\rm 2j,n}$,  to be $(GM_{\rm NS}/R_{\rm NS})/(0.5v^2_{\rm j,n})$ times larger. 
We checked one case with $v_{j,n}=10^5 \km \s^{-1}$ and found only small differences with respect to the case with $v_{j,n}=5 \times 10^4 \km \s^{-1}$.

\subsection{Simulations}
\label{subsec:Simulations}

We summarise the simulations we conduct in Table \ref{Table:cases}.
In all simulations the jets activity starts at $t=1\h$, and the properties of the ejecta are according to section \ref{subsec:flow}. Because the typical timescales of the simulations are hours we present the mass loss rate and power per hour. 
For numerical reasons we inject the jets at lower than the escape velocity from the NS. To have the same jets' power as in the real situation we increase the numerical mass outflow rate in the jets above that expected in the real situation.  
\begin{table}
\footnotesize
\centering
\begin{tabular}{|c|c|c|c|c|c|}
\hline
 Case  & $\Delta_{\rm cell,m}$ & $\dot E_{\rm 2j}$ & $\dot M_{\rm 2j,n}$ & $\alpha_j$ & Figs \\ 
   & $R_\odot$  & $\erg \h^{-1}$ & $M_\odot \h^{-1}$ &  & \\ 
 \hline 
LeN1 & $0.2344$ & $2.27 \times 10^{48}$ & $9.1 \times 10^{-5}$   & $20^\circ$  & 
\ref{fig:LeN1_dens_Evolution} - \ref{fig:PressureVel}\\ \hline
LeW2 & $0.2344$ & $2.27 \times 10^{48}$ & $9.1 \times 10^{-5}$   & $50^\circ$ & \ref{fig:PressureVel}\\ \hline
HeN3 & $0.2344$ & $8.46 \times 10^{48}$ & $3.4 \times 10^{-4}$   & $20^\circ$ & \ref{fig:PressureVel}\\ \hline
HeW4 & $0.2344$ & $8.46 \times 10^{48}$ & $3.4 \times 10^{-4}$   & $50^\circ$ & \ref{fig:PressureVel}\\ \hline
ExG5 & $0.4688$ & $2.27 \times 10^{48}$ & $9.1 \times 10^{-5}$   & $20^\circ$ & \ref{fig:ExG5} \\ \hline
\end{tabular}
\caption{Summary of the properties of the different simulations. In all cases we launch the jets at a velocity of $v_{j,n}=5 \times 10^4 \km \s^{-1}$.  From first to last column we present the name of the simulation, the smallest cell size in the grid, the power of the two jets, the mass outflow rate in the two jets, the jets' half opening angle, and the figures where we present results. In all cases we start to launch the jets at $t=60 \min$ after explosion. }
\label{Table:cases}
\end{table}
 
Between the different cases we simulate we vary the jets' power and the jets' half opening angle. 
We simulate cases with two jets' powers. The higher one is under the assumption that the NS accretes at the Bondi-Hoyle-Lyttleton rate and the jets carry $10 \%$ of the accretion energy (the jets carry $10 \%$ of the accreted mass at the escape velocity). The lower power is about quarter of the high power. 

The total energy of the ejecta in our simulations is $E_{\rm SN}=10^{51} \erg$. However, we should compare the energy of the jets with the energy of the ejecta that the jets interact with. Consider then the energy of the ejecta with velocity $v< v_{\rm ej,i}$. From the density and velocity profiles (section \ref{subsec:flow}) the ejecta energy in the relevant range of $v< v_{\rm ej,i}$ is 
\begin{eqnarray}
\begin{aligned}
& E_{\rm ej} (<v_{\rm ej,i}) = \int^{v_{\rm ej,i} t}_0 \rho_0 \left( \frac{r}{t v_{\rm br}} \right)^{-1} \frac{1}{2} v^2 4 \pi r^2 dr 
\\ &
= \frac{5}{9} E_{\rm SN}  
\left( \frac{v_{\rm ej,i}} {v_{\rm br}} \right)^4  
 = 1.7 \times 10^{-3}    
\left( \frac{v_{\rm ej,i}}{0.235 v_{\rm br}} \right)^4 E_{\rm SN}.
\end{aligned}
\label{eq:Eej(Vejd)}
\end{eqnarray}
The velocity of the ejecta that reaches the NS after an hour is $5 R_\odot / 1 h = 970 \km \s^{-1}$. Later we find that the jets inflate a bubble that expands faster than the ejecta and reaches to regions where the ejecta velocity is $v_{\rm ej,i} \simeq 3000-3500 \km \s^{-1}$. The corresponding ejecta energy is 
$E_{\rm ej} (<3000 \km \s^{-1}) = 2.7 \times 10^{49}  \erg$. The jets inflate the bubble into one side, so the relevant energy of the ejecta is $\approx 10^{49} \erg$.
Indeed, the jets' energy that we inject within 2h is about equal to this energy. We present the consequences in section \ref{sec:Results}.

\section{Numerical set up}
\label{sec:Numerical}

We use version 4.2.2 of the adaptive-mesh refinement (AMR) hydrodynamical FLASH code \citep{Fryxell2000} in its 3D mode. 
We employ a full 3D AMR using a Cartesian grid $(x,y,z)$ with outflow boundary conditions at all boundary surfaces. The origin of the grid is at the centre of the ejecta (the progenitor of the CCSN). The $z=0$ plane is the equatorial plane of the flow. We simulate the whole space (the two sides of the equatorial plane). 

As the strong jet-ejecta interaction takes place in optically thick regions, we turn off radiative cooling at any gas temperature.
The equation of state includes both radiation pressure and gas pressure with an adiabatic index of $\gamma=5/3$, due both to ions and electrons, i.e., $P_{\rm tot} = P_{\rm rad} + P_{\rm ion}+P_{\rm elec}$.

We use resolution with 7 refinement levels. 
In four cases the minimum cell size is $\Delta_{\rm cell,m}=1.64 \times 10^{10} \cm = 0.234 R_\odot$, and the total size of the Cartesian numerical grid is $(120 R_\odot)^3$, i.e., $(L_x,L_y,L_z) = \pm 60 R_\odot$.
In one case the cells are twice as large and so is the numerical grid.   

We simulate a CCSN ejecta with a mass of $M_{\rm ej}=3.5 M_\odot$ and a kinetic energy if $E_{\rm SN}=10^{51} \erg$. 
We take the ejecta at $t=1\h$ after the explosion, such that the initial (when we start the simulation) velocity at each radius is $v(r)=r/1 \h$, and the density is according to equation (\ref{eq:rho(rt)}) at $t=1\h$. 

We inject the two opposite jets along a constant axis which is parallel to the $z$ axis at $(x,y,z)=(5R_\odot,0,0)$. We inject each jet in a cone with a half opening angle of $\alpha_{\rm j} = 20^\circ$ or $50^\circ$. The length of each injection-cone is $\Delta r_{\rm j}=R_\odot$ (or $\Delta r_{\rm j}=2 R_\odot$ for the low resolution case). 
At the beginning of the simulation the two opposite cones are filled with the jets material. 

For numerical reasons (to avoid very low densities) we inject a very weak slow wind in the directions where we do not launch the jets, i.e., in the sector $\alpha_{\rm j} < \theta \le 90^\circ$ in each hemisphere. Because of the constant-density sphere near the center, and the region where we numerically inject the jets, the flow structure close to the center includes some numerical effects.
The initial temperature of the simulation box and the jets is $10000\K$.

\section{Results}
\label{sec:Results}

Our goal is to present the general inflation process of a bubble by jets that a NS companion launches, and to find its region of influence within the ejecta. We conduct simulations with the parameters that we present in Table \ref{Table:cases}. We present the evolution only for Case LeN1 (section \ref{subsec:LeN1}) and for the Case ExG5 that has the same parameters but with an extended grid (section \ref{subsec:ExG5}). For the other cases we present their pressure and velocity maps at $t=150 \min$ after explosion (section \ref{subsec:OtherCases}). In all cases we start the simulation at $t=1 \h$ when jets start to be active. 

\subsection{Case LeN1}
\label{subsec:LeN1}
\begin{figure} 
\centering
\begin{tabular}{cc}
\includegraphics[trim=0.0cm 6.0cm 0cm 5.2cm ,clip, scale=0.39]{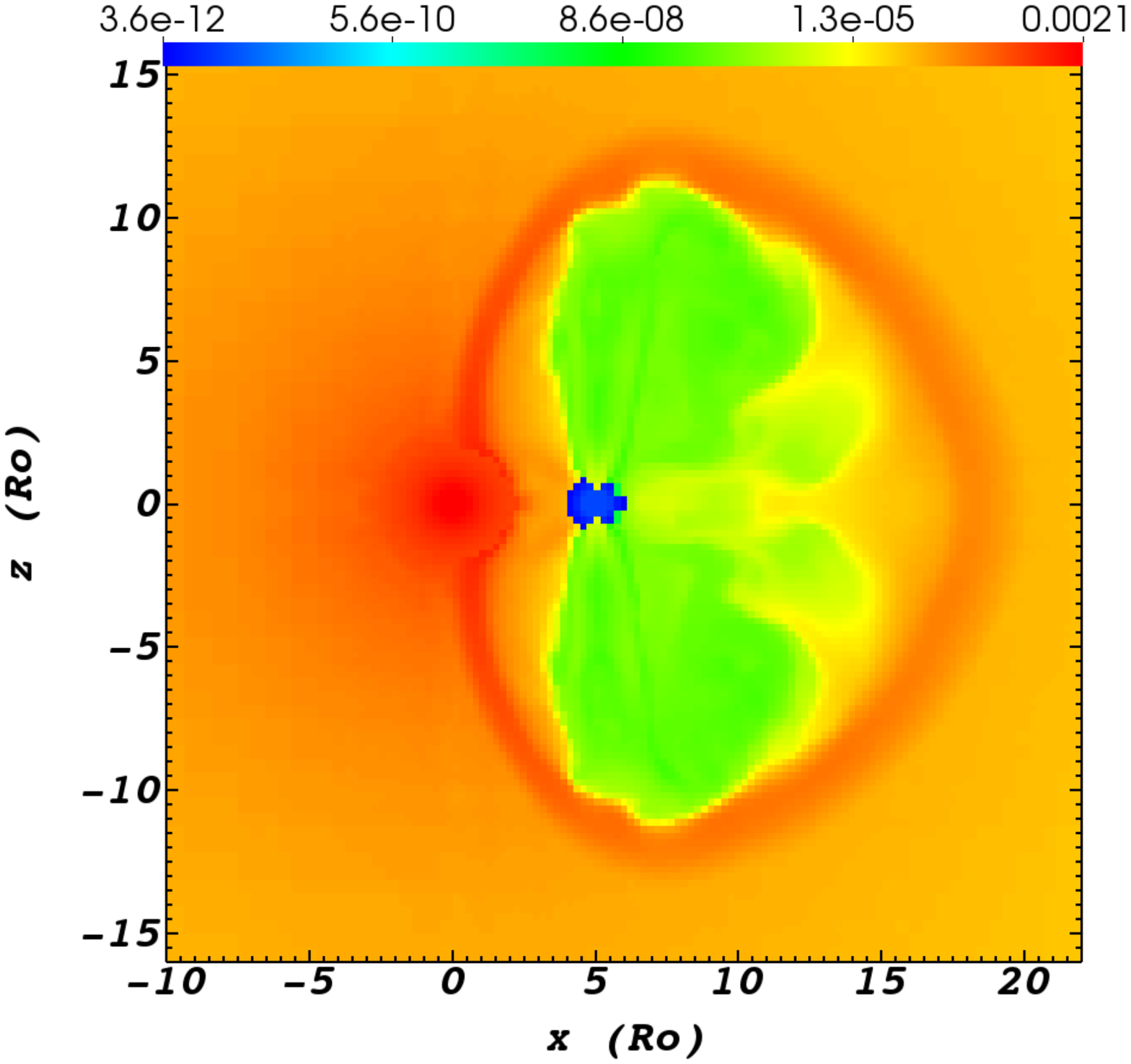} \\
\includegraphics[trim=0.0cm 6.0cm 0cm 5.2cm ,clip, scale=0.39]{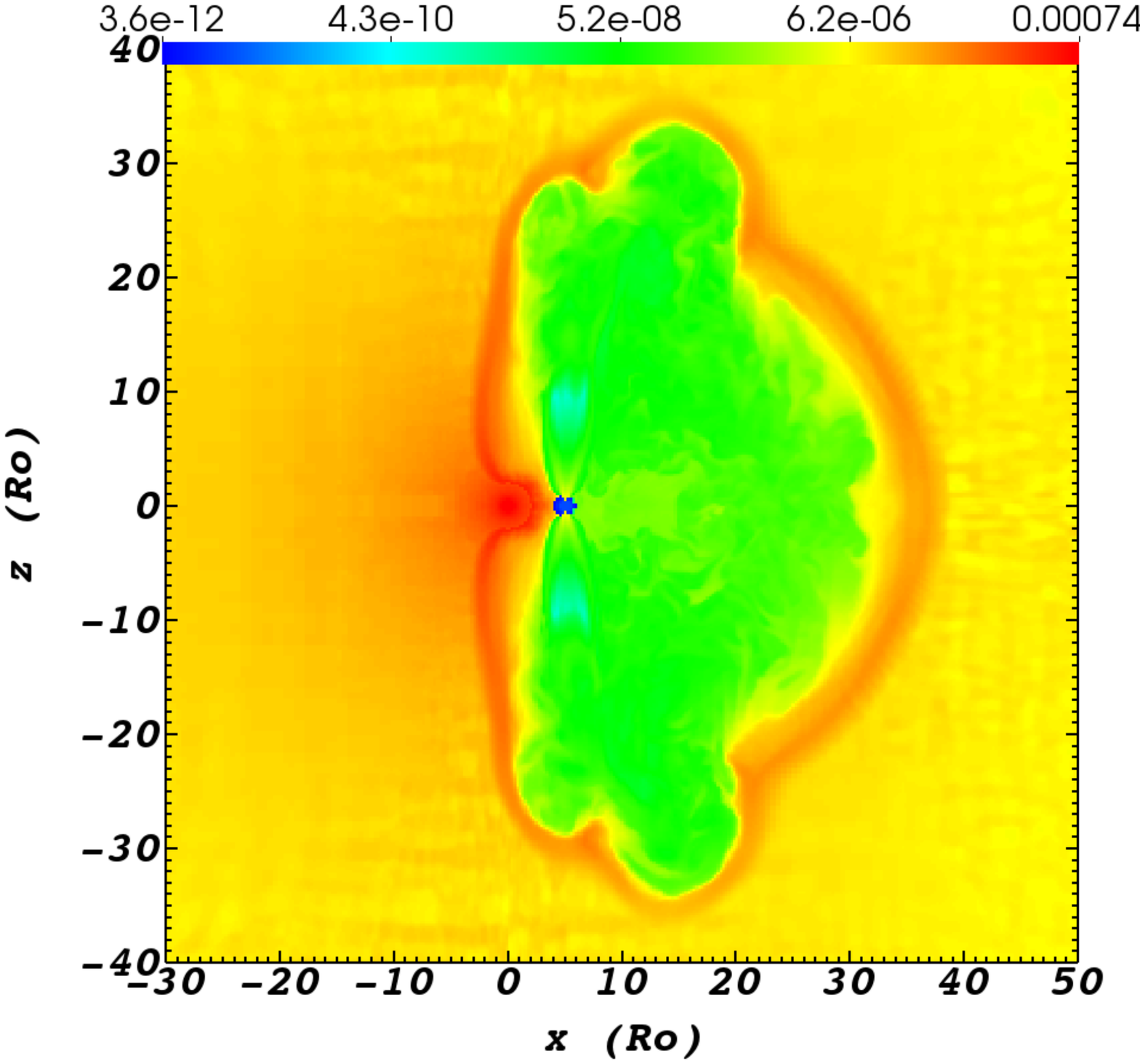}\\
\includegraphics[trim=0.0cm 6.0cm 0cm 5.2cm ,clip, scale=0.39]{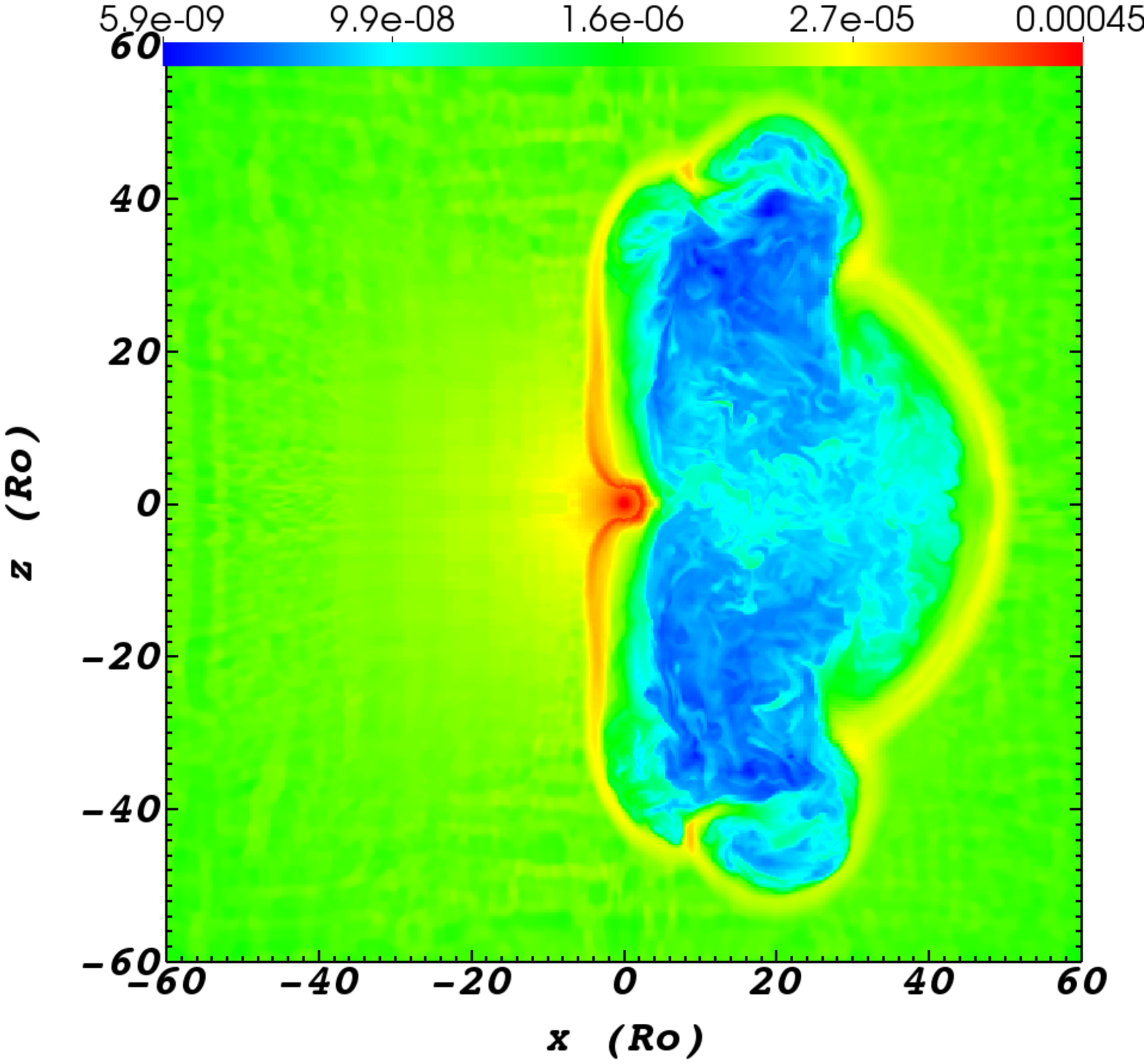} 
\end{tabular}
\caption{Density maps in the meridional plane of Case LeN1 (details of the cases in Table \ref{Table:cases}) at three times, from top to bottom, $t=90 \min$,  $t=150 \min$, and $t=192 \min$ after explosion. The density scale is according to the color bar in units of $\g \cm^{-3}$. Note the increasing size of the panels with time and the different density scales of the color-bars. 
}
  \label{fig:LeN1_dens_Evolution}
    \end{figure}

\begin{figure} 
\centering
\begin{tabular}{cc}
\includegraphics[trim=0.0cm 6.0cm 0cm 5.2cm ,clip, scale=0.39]{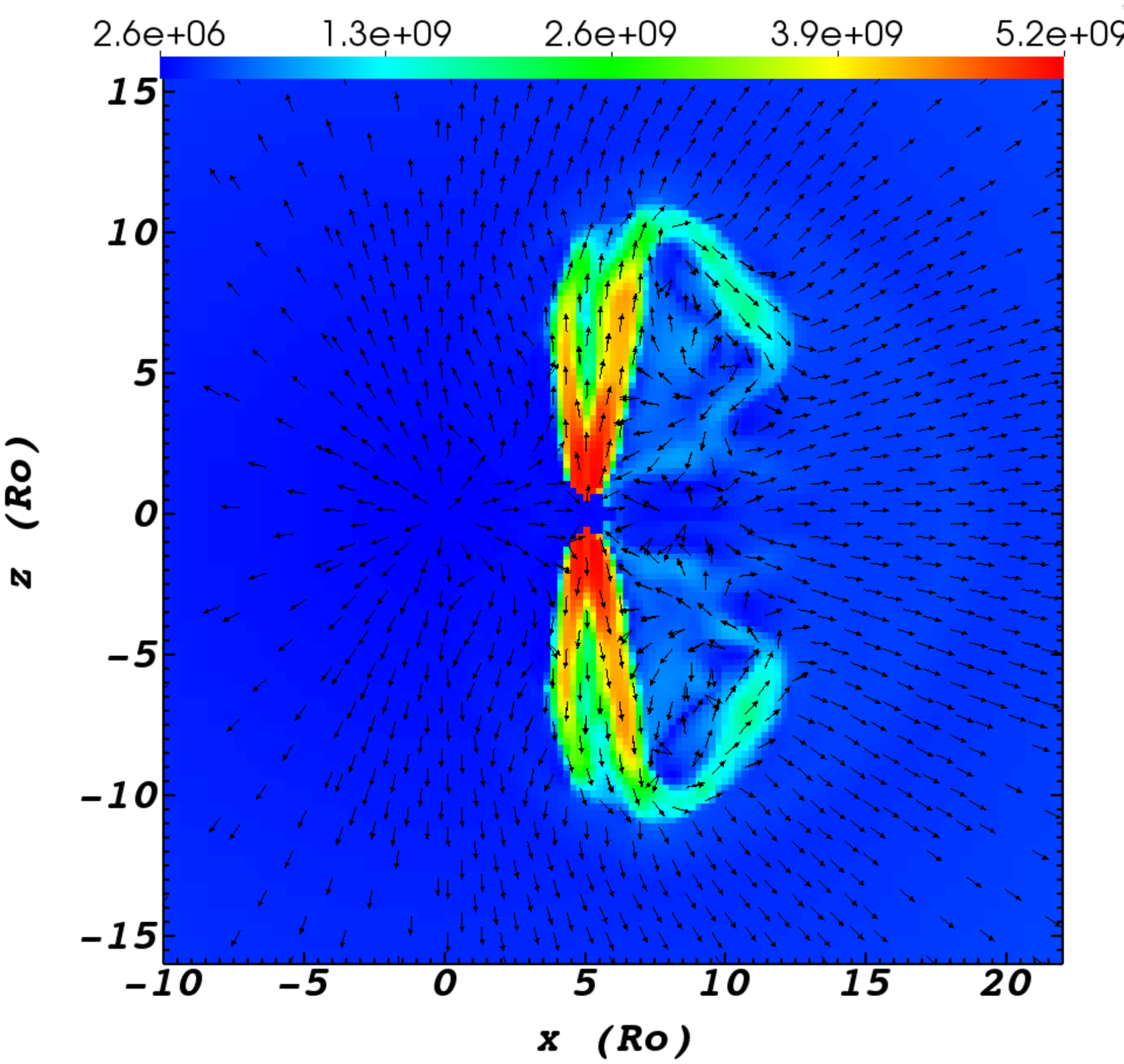} \\
\includegraphics[trim=0.0cm 6.0cm 0cm 5.2cm ,clip, scale=0.39]{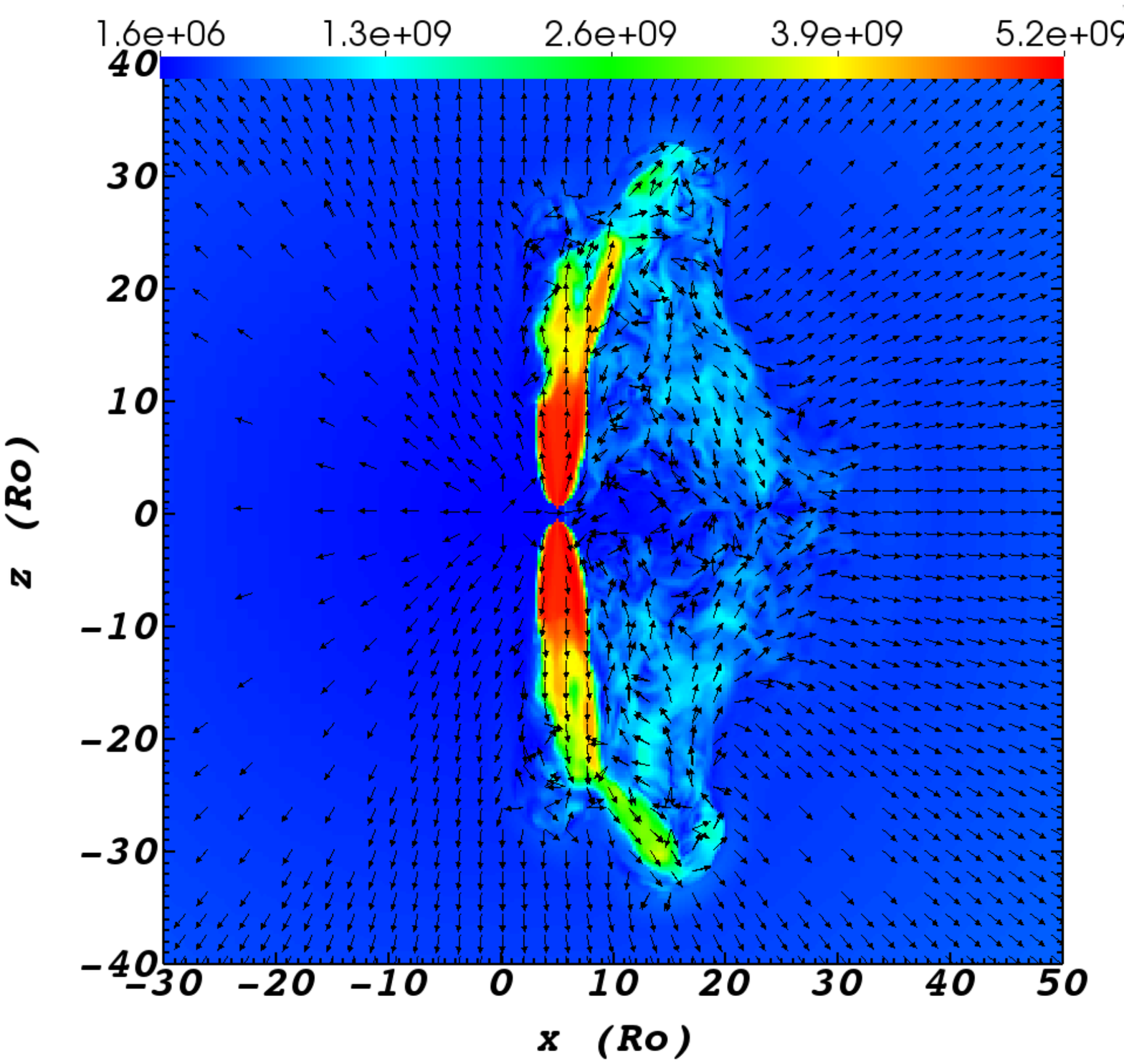}\\
\includegraphics[trim=0.0cm 6.0cm 0cm 5.2cm ,clip, scale=0.39]{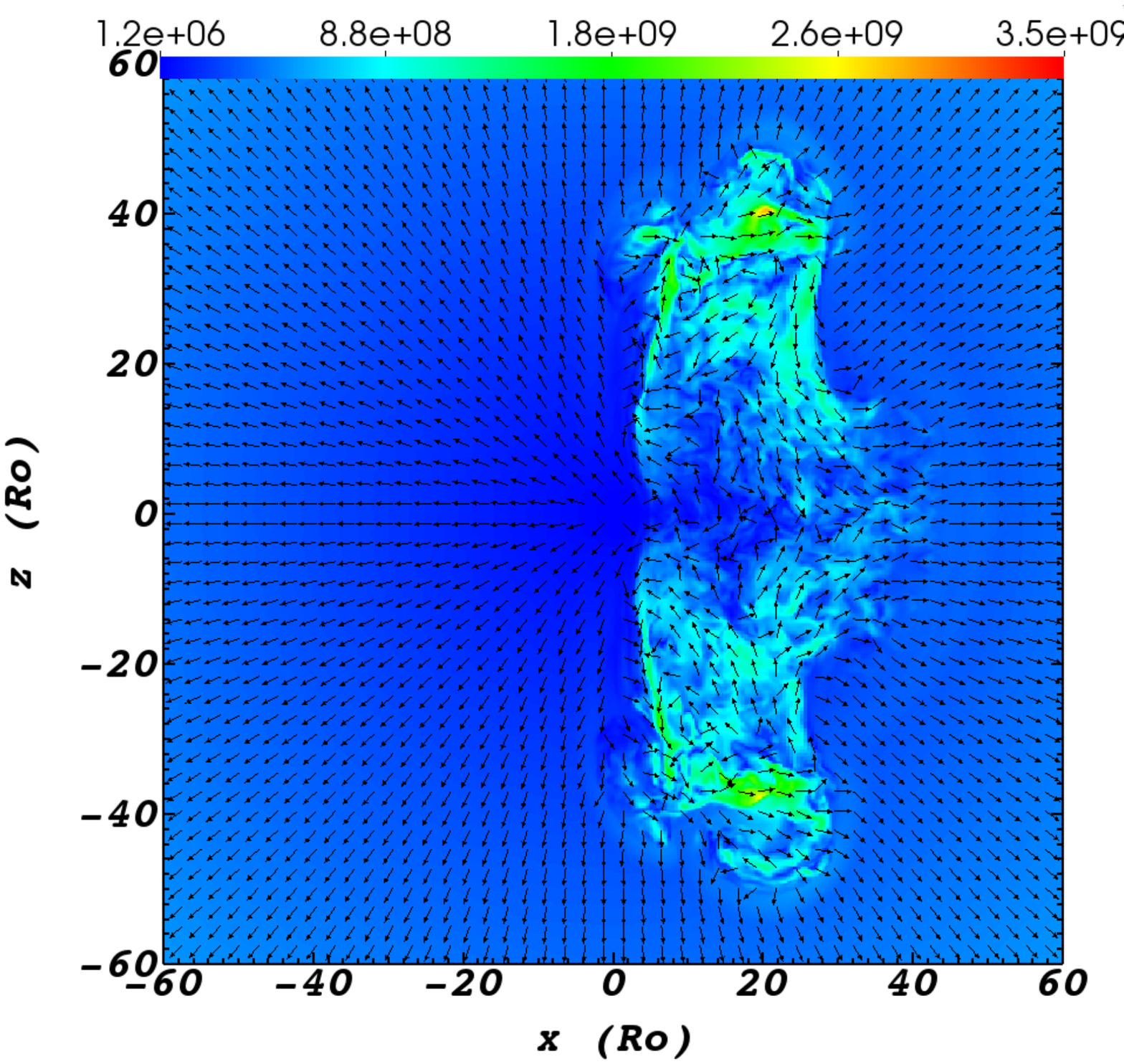} 
\end{tabular}
\caption{Velocity maps in the meridional plane of the  simulation at three times as 
in Fig. \ref{fig:LeN1_dens_Evolution}.
The velocity scales are according to the color bars in units of $\cm \s^{-1}$ (all arrows have the same length). Note the increasing size of the panels with time and the different velocity scales of the color-bars. }
  \label{fig:LeN1_vel_Evolution}
    \end{figure}

\begin{figure} 
\centering
\begin{tabular}{cc}
\includegraphics[trim=0.0cm 6.0cm 0cm 5.2cm ,clip, scale=0.39]{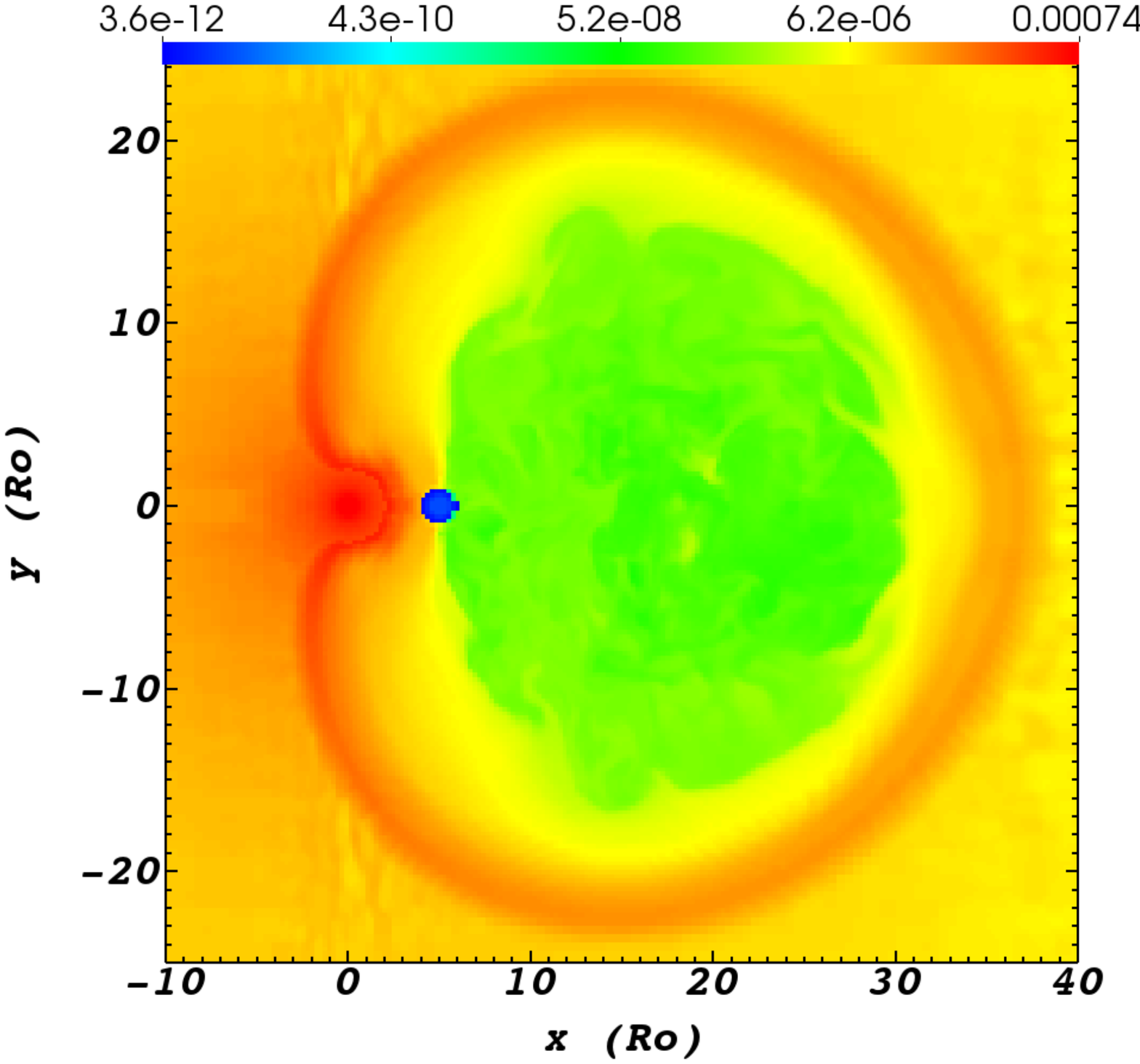} \\
\includegraphics[trim=0.0cm 6.0cm 0cm 5.2cm ,clip, scale=0.39]{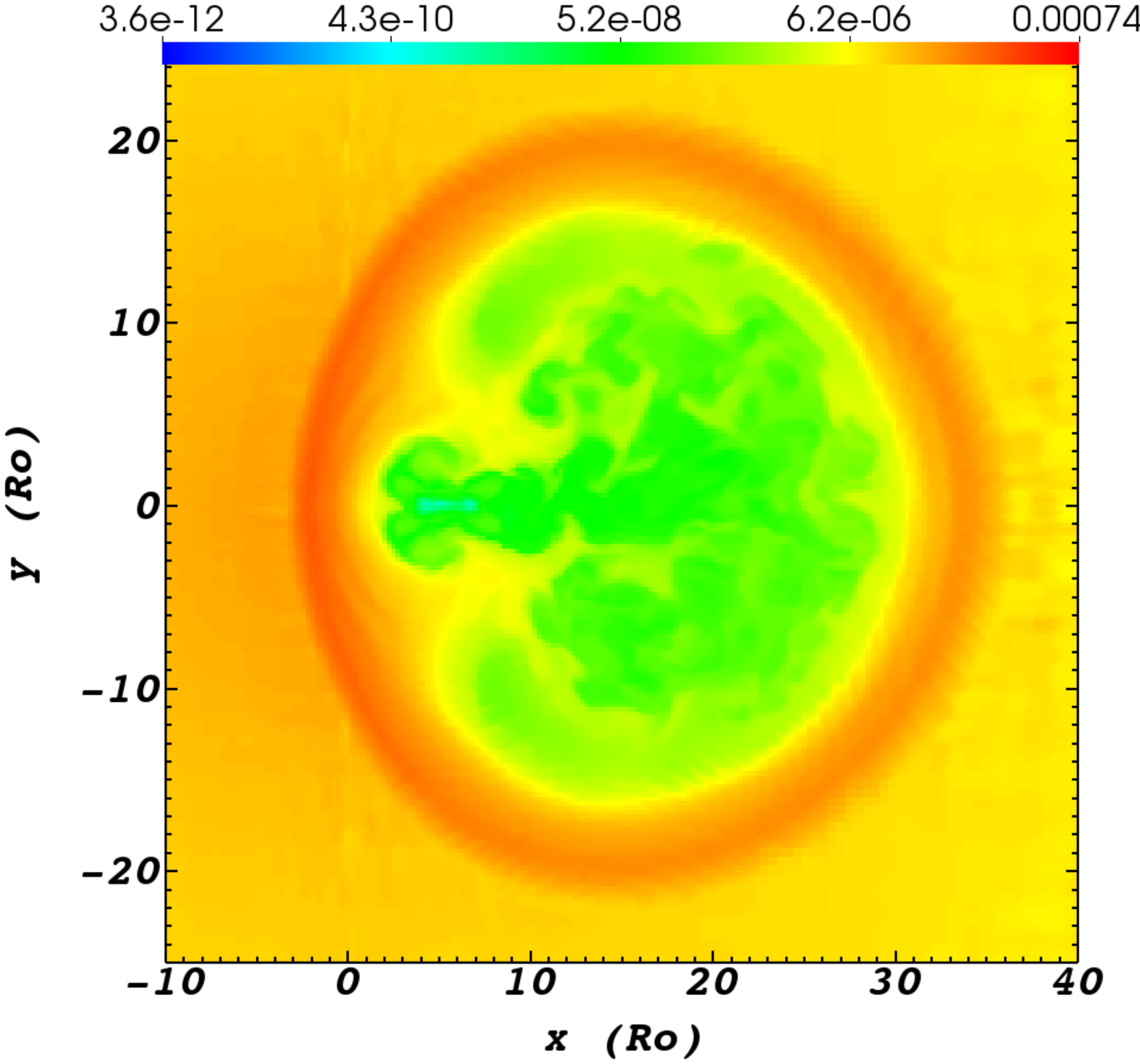}\\
\includegraphics[trim=0.0cm 6.0cm 0cm 5.2cm ,clip, scale=0.39]{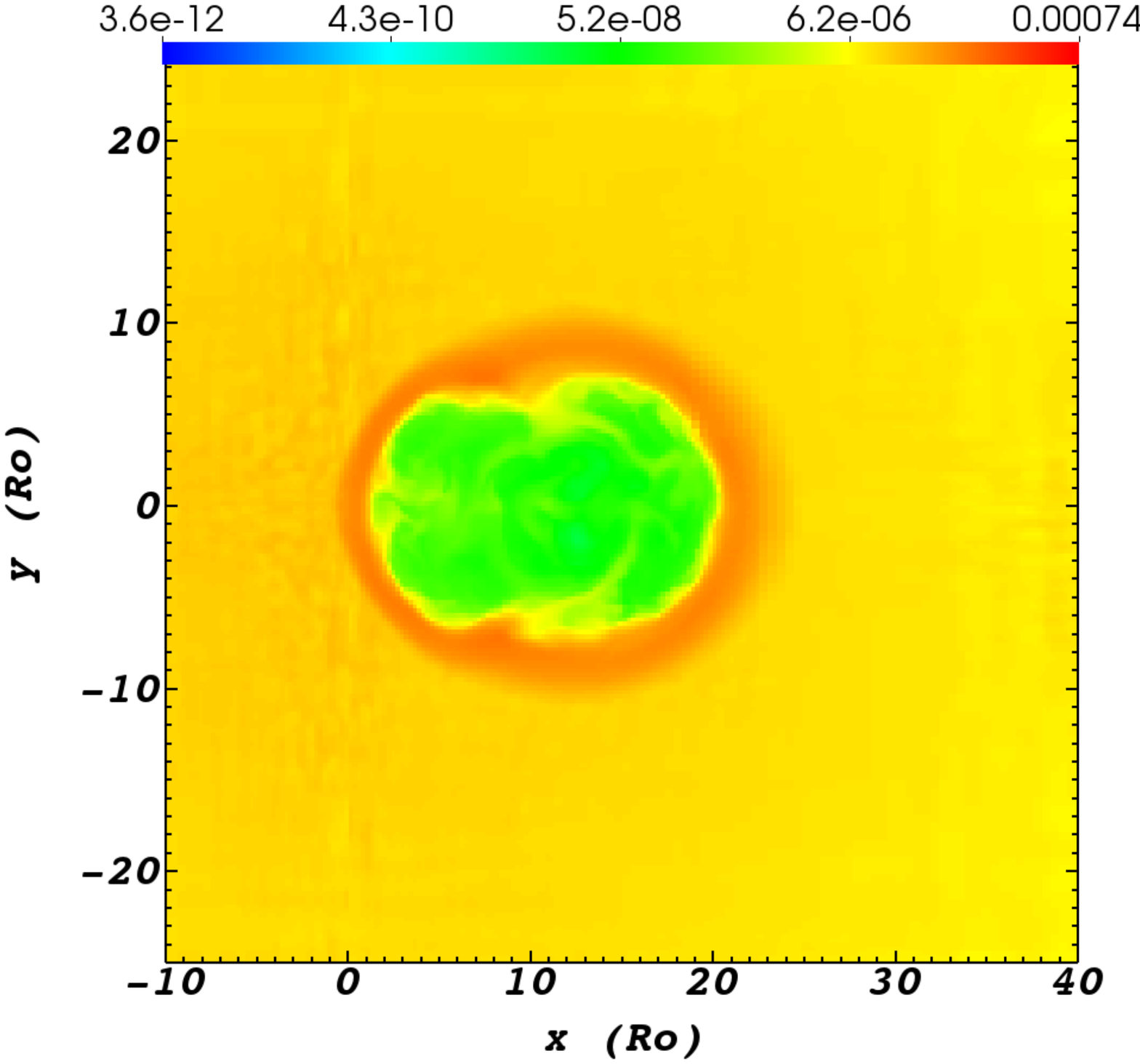} 
\end{tabular}
\caption{Density maps of the Case LeN1 at $t=150 \min$ after explosion in the  planes (from top to bottom): $z=0$, $z=10 R_\odot$, and $z=25 R_\odot$. The density scale is according to the color bar in units of $\g \cm^{-3}$. }
  \label{fig:LeN1_densXY_Evolution}
    \end{figure}

In Fig. \ref{fig:LeN1_dens_Evolution} we present the density maps of Case LeN1 in the meridional plane that contains the NS companion and the center of explosion at three times (note the different scales of the panels, both in the axes sizes and in the color-bar scales).
The NS companion that launches the jets is at $(x,y,z)=(5 R_\odot,0,0)$, and the initial symmetry axis of the jets is along the $z$ axis (perpendicular to the equatorial plane $z=0$). In Fig. \ref{fig:LeN1_vel_Evolution} we present the velocity maps in the same plane and at the same times. In Fig. \ref{fig:LeN1_densXY_Evolution} we present the density in the planes $z=0$, $z=10R_\odot$ and $z=25 R_\odot$ at $t=150 \min$. The blue circle in the upper panel is the very low density of the freely expanding jets.  

\begin{figure} 
\centering
\begin{tabular}{cc}
\includegraphics[trim=0.0cm 6.0cm 0cm 5.2cm ,clip, scale=0.39]{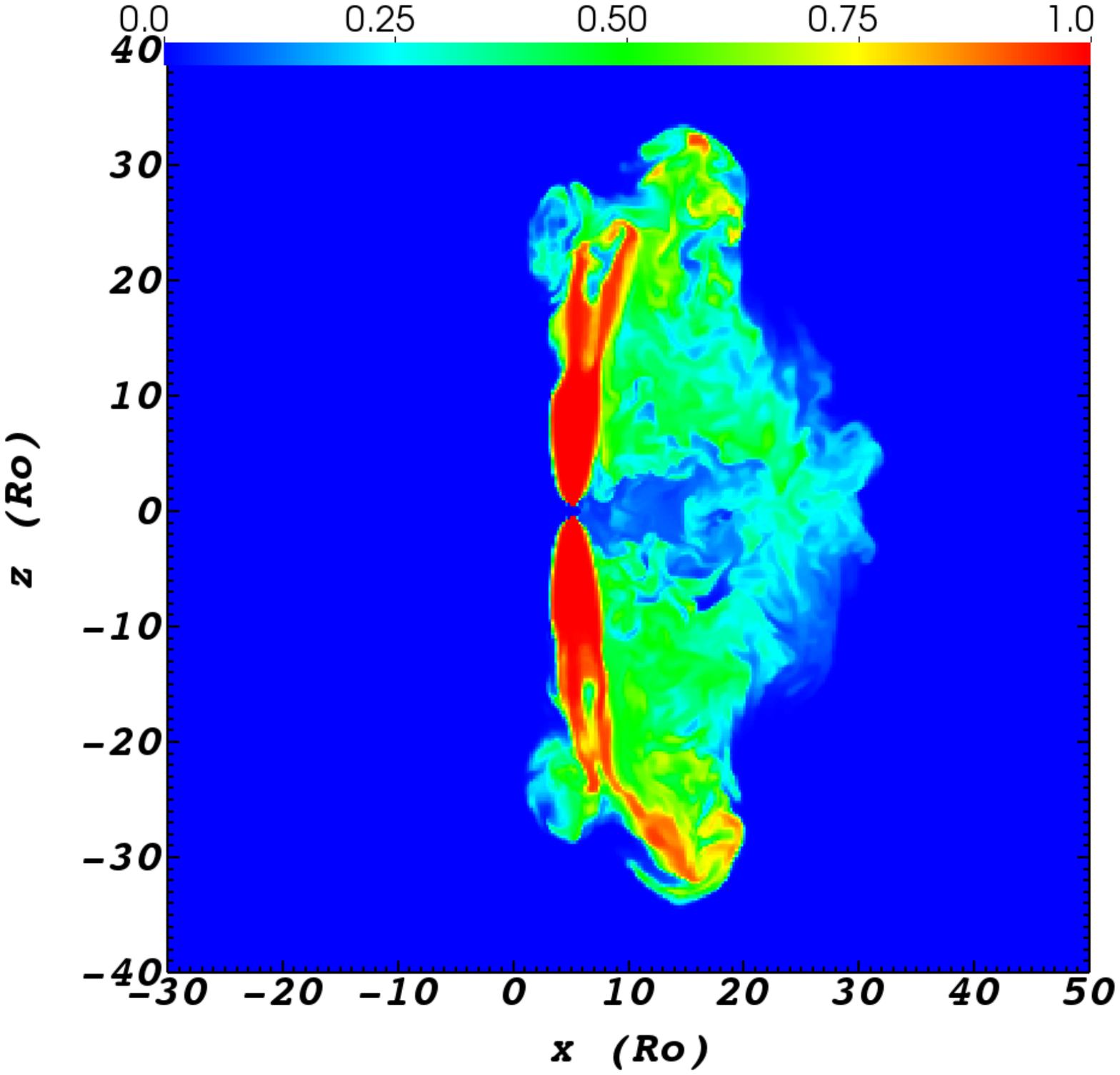} \\
\includegraphics[trim=0.0cm 6.0cm 0cm 5.2cm ,clip, scale=0.39]{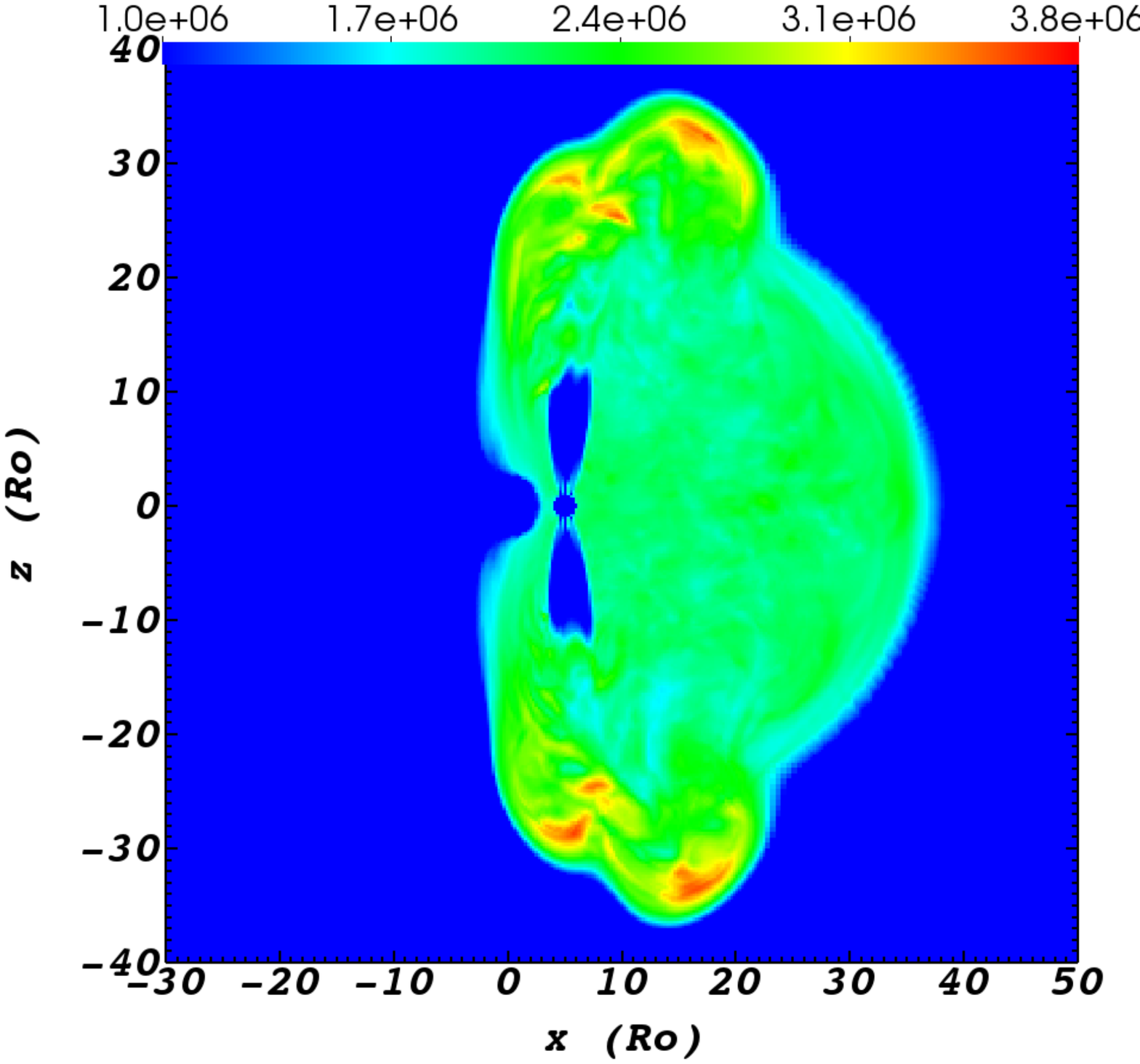}\\
\includegraphics[trim=0.0cm 6.0cm 0cm 5.2cm ,clip, scale=0.39]{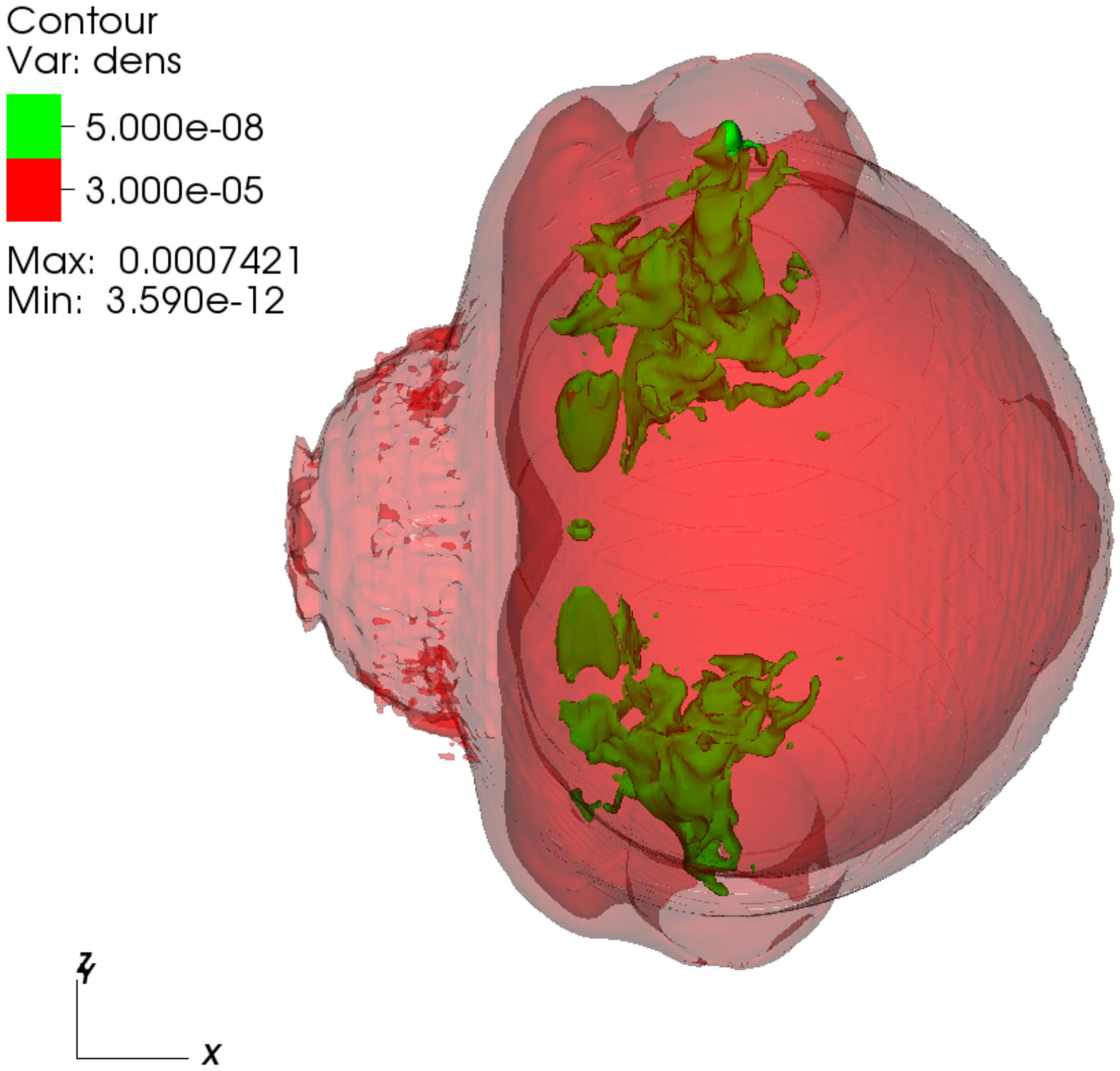} 
\end{tabular}
\caption{Some properties of the the Case LeN1 at $t=150 \min$. In the upper and middle panels  we present the jet-tracer map and temperature (temperature in K by the color bar), respectively, in the meridional plane. In the lower panel we present a three-dimensional view of the density (green for $\rho=5\times 10^{-8} \g\ cm^{-3}$ and red for $\rho = 3 \times 10^{-5} \g\ cm^{-3}$). The observer is in the direction of $(x,y,z,)=(0,1,1)$.  }
  \label{fig:LeN1_3D_temp_tracer}
    \end{figure}

In Fig. \ref{fig:LeN1_3D_temp_tracer} we present three different panels all at $t=150 \min$. In the upper panel we present the jet-tracer map in the meridional plane as in Figs.  
\ref{fig:LeN1_dens_Evolution} and Fig. \ref{fig:LeN1_vel_Evolution}.  
The jet-tracer variable follows the gas that was injected in the jets, and in each grid cell it is equal to the fraction of the gas that originated in the jets.
In the second panel we present the temperature map in the same plane. 
In the lower panel we present a three-dimensional view by two colors. The green color depicts the low density ($\rho=5\times 10^{-8} \g\ cm^{-3}$) regions of the post-shock jets' material, while the red color depicts higher density gas of $\rho = 3 \times 10^{-5} \g\ cm^{-3}$ in the bubble-ejecta boundary. The view of the 3D density map is from the direction $(x,y,z)=(0,1,1)$
  
We turn to analyse the results. 

\subsubsection{The general morphology}
\label{subsubsec:Morphology}
At a very early time (not shown in the figures) the two bubbles, one from each jet, merge to form one bubble. The bubble is the green volume in the two upper panels of Fig. \ref{fig:LeN1_dens_Evolution}, which is bounded by a dense shell (red color in these panels). The material inside the dense shell, the bubble, includes shocked ejecta gas (closer to the shell boundary on the right side of the bubble of Fig. \ref{fig:LeN1_dens_Evolution}) and shocked jets' gas.
Namely, there is a forward shock running into the ejecta. Behind this shock is the dense shell. The jets pass through the reverse shocks, one at each side of the equatorial plane, that we see by the sharp jump in temperature in the middle panel of Fig. \ref{fig:LeN1_3D_temp_tracer} (blue to green inside the bubble). The upper panel of fig. \ref{fig:LeN1_3D_temp_tracer} presents the jet-tracer map. Comparing the tracer map with the temperature map (middle panel of fig. \ref{fig:LeN1_3D_temp_tracer}) teaches us that the high temperature region contains indeed both post-shock jets' gas and post-shock ejecta gas. 

The CCSN ejecta pushes the bubble away from the center (to the right in Figs. \ref{fig:LeN1_dens_Evolution} and \ref{fig:LeN1_vel_Evolution}). During the two hours activity period of the jets, the two jet termination shocks move to larger distances from the equatorial plane (two upper panels of Fig. \ref{fig:LeN1_vel_Evolution}). The post-shock jets' material flows toward the equator in a back-flow stream at larger distances from the center relative to the original jets. This meridional flow has a large-scale mirror symmetry about the equatorial plane. 
  
The density color-coding changes between the three panels of Fig. \ref{fig:LeN1_dens_Evolution}. In the upper two panels the blue color presents the very low density of the freely expanding jets. In the lower panel, where the jets are not active anymore, the blue color presents the very low-density regions of the bubble. The density inside the bubble is $\approx 0.01$ times that of the ejecta. Due to the high temperature (middle panel of Fig. \ref{fig:LeN1_3D_temp_tracer}) radiation pressure dominates inside the bubble, and the pressure in the bubble is larger than in the ejecta (section \ref{subsec:OtherCases}). The lowest density regions inside the bubble (deep blue in the lower panel of Fig.  \ref{fig:LeN1_dens_Evolution}) show that each jet inflated a different very-low density region, but the separate bubbles merged at a very early time to form one bubble.  

At a time of $t=1 \h$ when we start the jets' activity the ejecta that crosses the NS has a velocity of $5 R_\odot / 1 \h=967 \km \s^{-1}$. By the time we end the simulation here (because the bubble reaches the edge of the  numerical grid) the dense shell boundary of the bubble reaches ejecta material with  velocities of $\simeq 3000 \km \s^{-1}$. We explore the bubble evolution at later times with a larger low-resolution grid in section \ref{subsec:ExG5}.

Fig. \ref{fig:LeN1_densXY_Evolution} shows that the cross section of the bubble narrows down with increasing  distance from the equatorial plane.   

\subsubsection{Instabilities and vortexes}
\label{subsubsec:Instability}

The flow of the  post-shock jets' material, namely, the flow inside the bubble, has a complicated structure (Fig. \ref{fig:LeN1_vel_Evolution}). The flow includes a large-scale meridional back-flow on each side of the equatorial plane, small vortexes, and a radial outflow on the outer boundary of the bubble. We will present the velocity maps of the other cases in section \ref{subsec:OtherCases}.  

The pressure in the bubble is larger that that of he ejecta and the density inside the bubble is lower than that of the ejecta. Therefore, the bubble-ejecta boundary is Rayleigh–Taylor unstable. Indeed, the upper panel of Fig. \ref{fig:LeN1_dens_Evolution} clearly shows high-density instability tongues that protrude from the dense shell into the low density bubble. In the middle panel these tongues are smaller. The reason is that the strong meridional flow (Fig. \ref{fig:LeN1_vel_Evolution}) washes out the tongues. 
The relatively high density clumps and filaments (yellow) inside the bubble (green) in the three panels of Fig. \ref{fig:LeN1_densXY_Evolution} also demonstrate the instabilities in the bubble-ejecta boundary.

\subsubsection{Mixing}
\label{subsubsec:Mixing}
The instabilities and vortexes mix post-shock ejecta gas with post-shock jets' gas. The upper panel of Fig. \ref{fig:LeN1_3D_temp_tracer} presents the degree of mixing between the media. The green areas in this jet-tracer map show that a large volume inside the bubble has about equal fraction of jet-originated and ejecta gases. The inner parts of the ejecta contain newly synthesised isotopes (e.g., \citealt{Wongwathanaratetal2017}), and the bubble can change the flow structure of these isotopes in the side of the ejecta that the jets influence.    
    
\subsection{Other cases}
\label{subsec:OtherCases}
 
\begin{figure*} 
\centering
\begin{tabular}{cc}
\includegraphics[trim=1.4cm 6.0cm 0cm 5.2cm ,clip, scale=0.45]{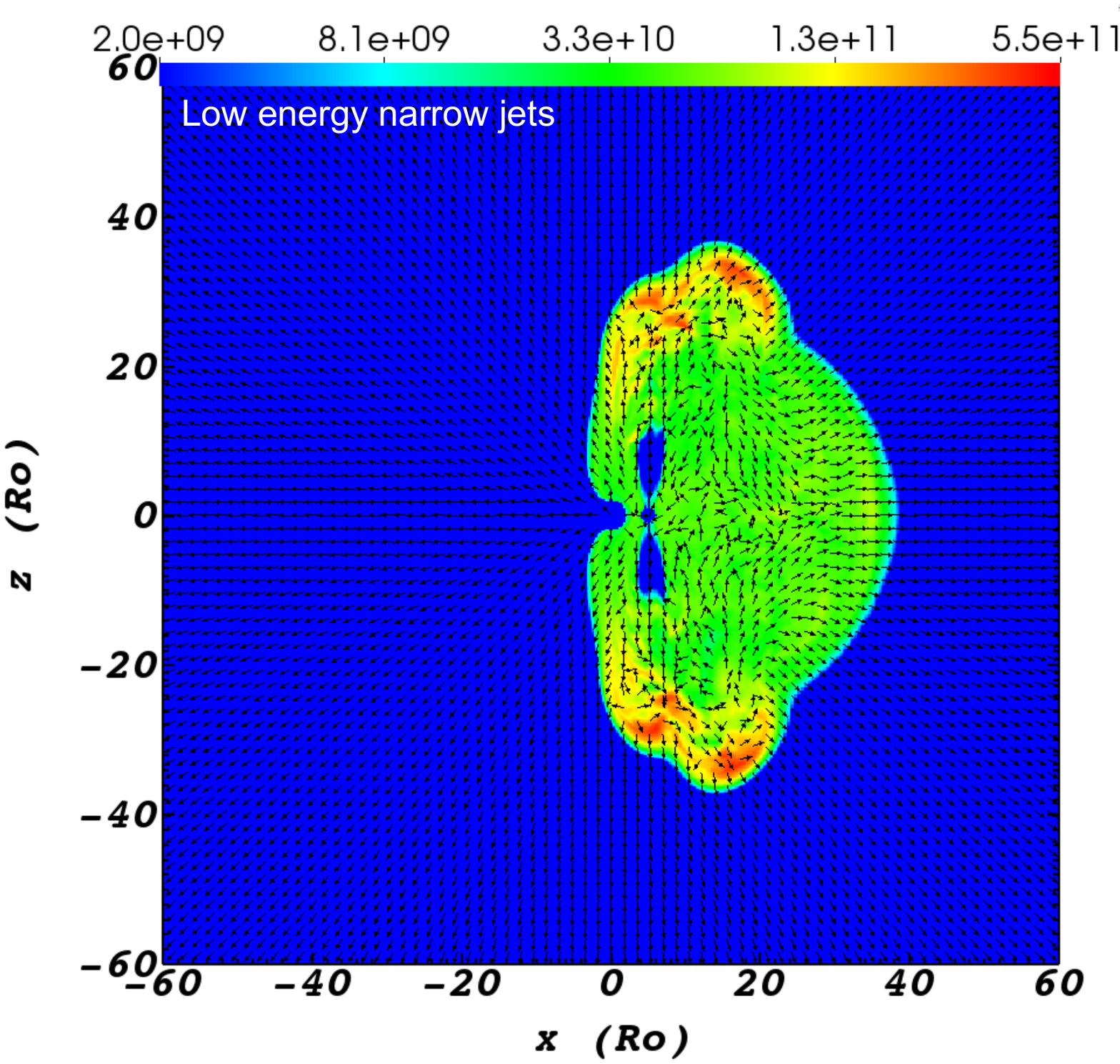} 
\includegraphics[trim=1.4cm 6.0cm 0cm 5.2cm ,clip, scale=0.45]{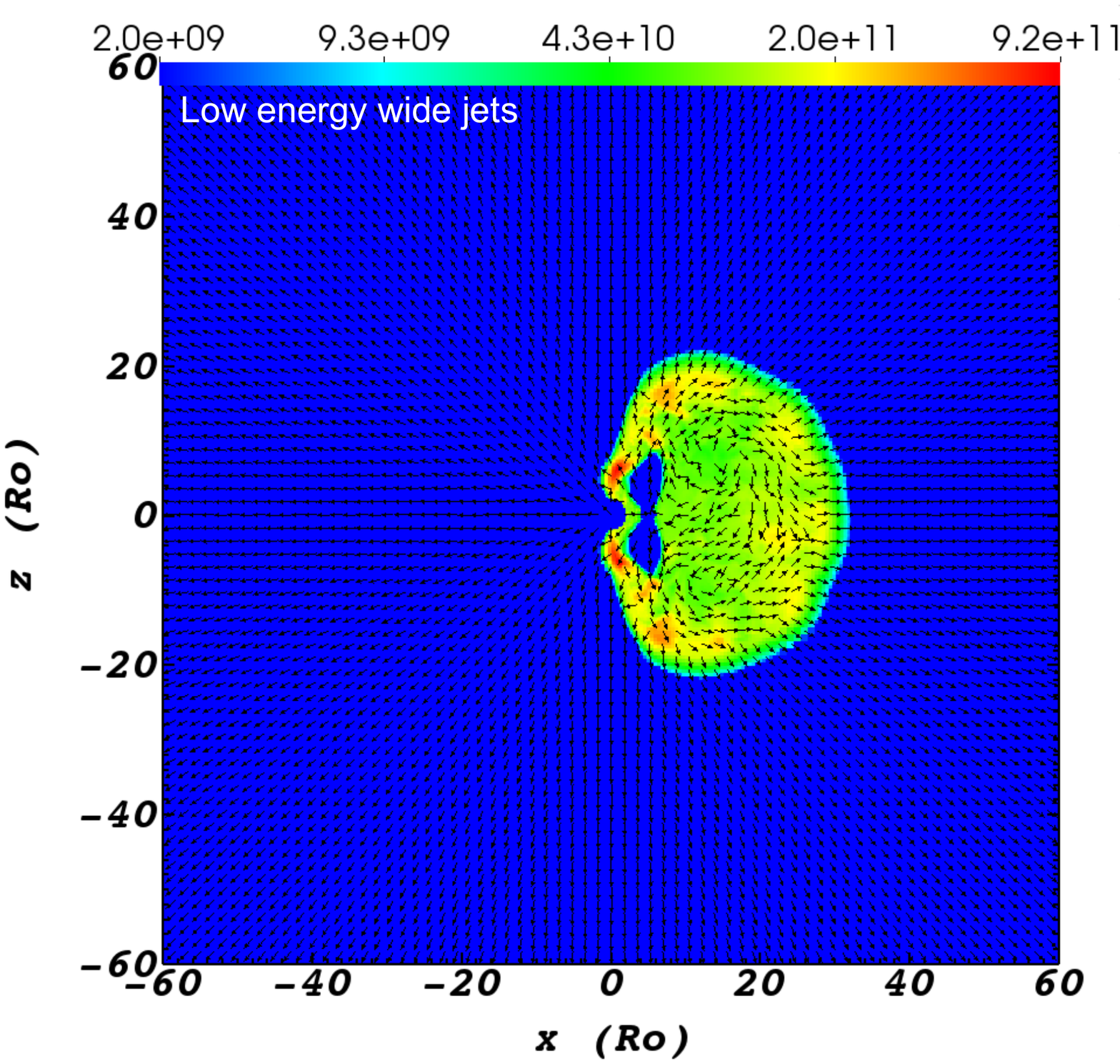}\\
\includegraphics[trim=1.4cm 6.0cm 0cm 5.2cm ,clip, scale=0.45]{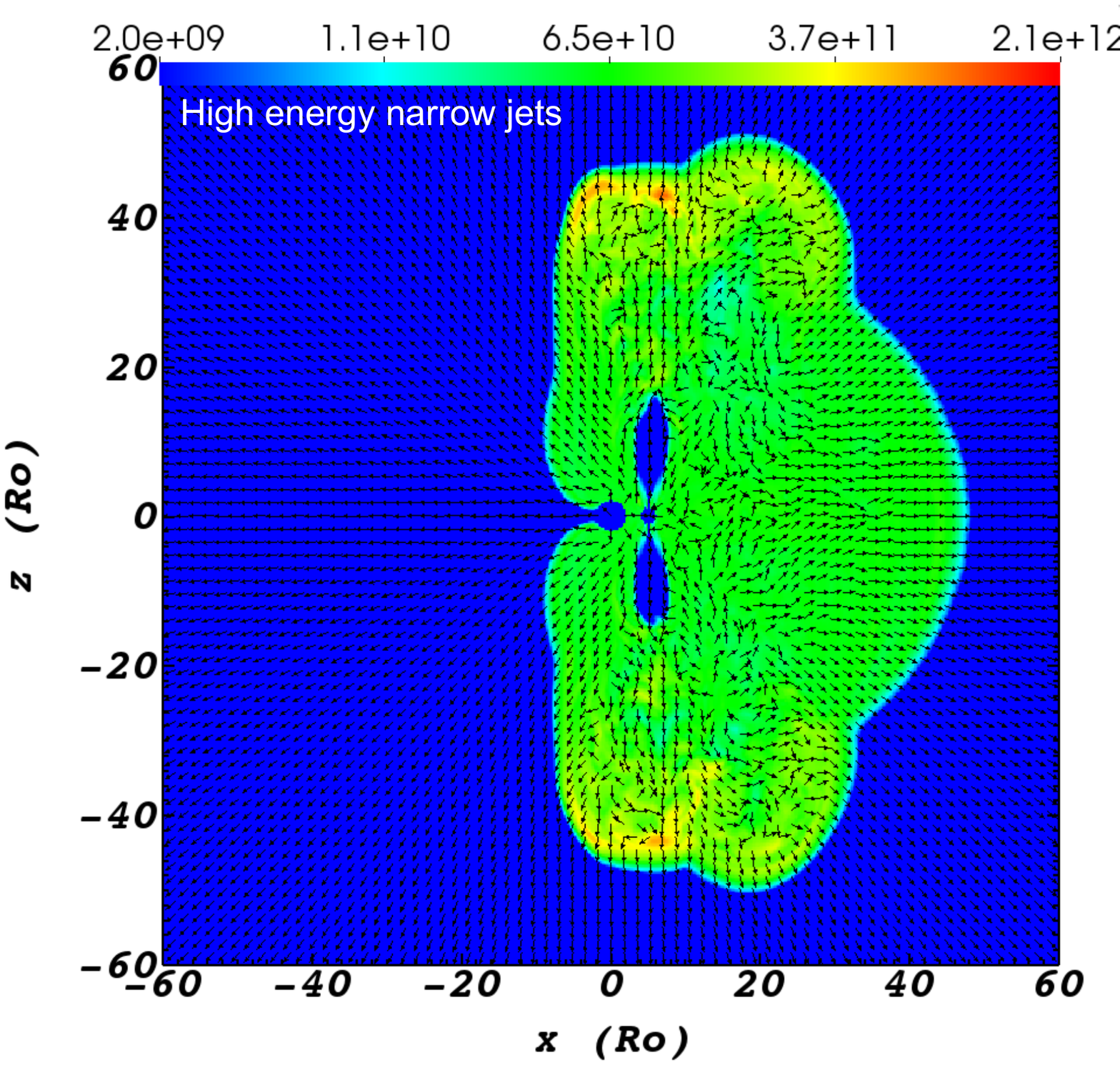} 
\includegraphics[trim=1.4cm 6.0cm 0cm 5.2cm ,clip, scale=0.45]{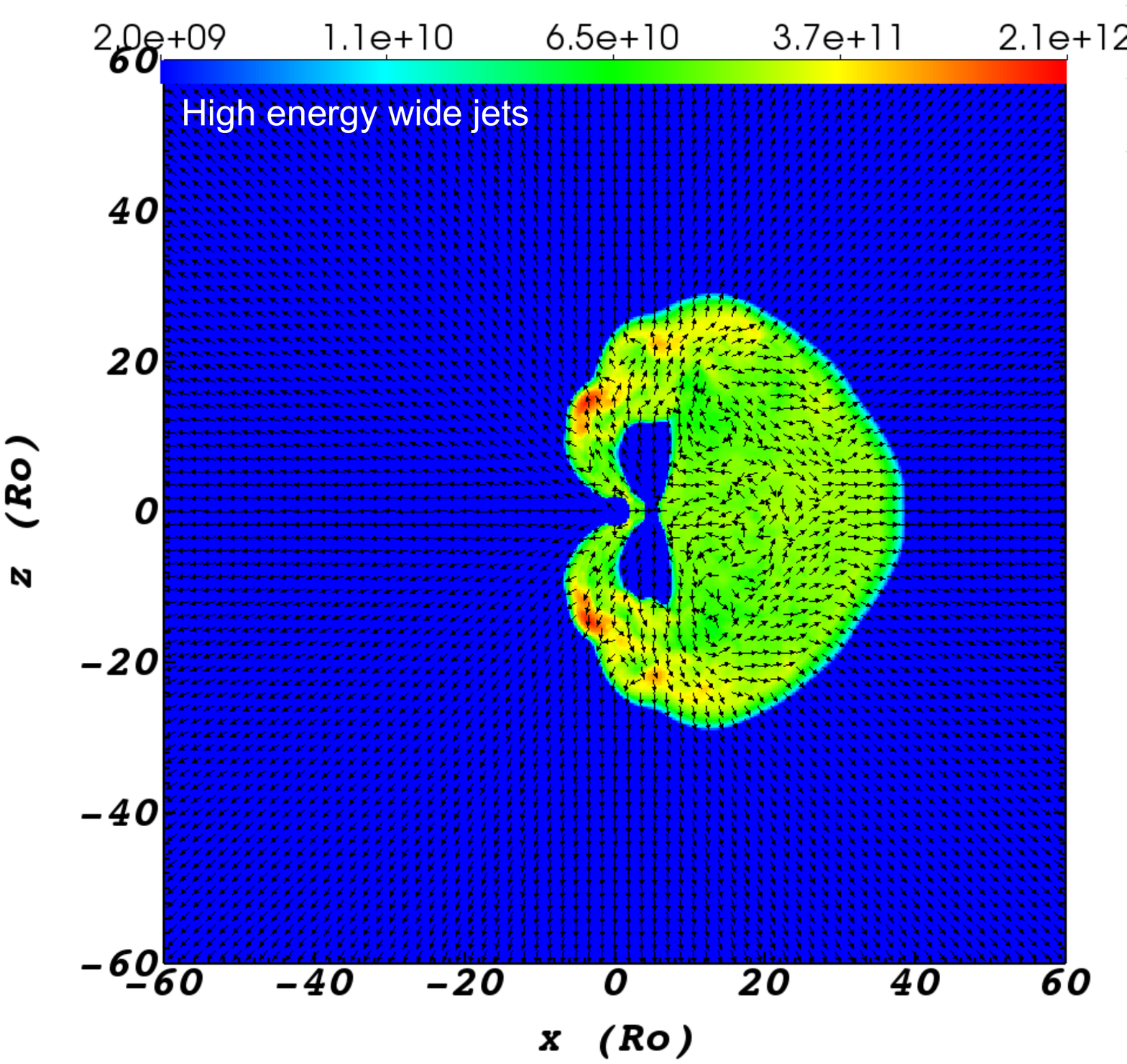} 
\end{tabular}
\caption{Pressure maps and stream maps (where arrows present only the direction of the flow) in the meridional plane that contains the NS and the center of explosion at $t=150 \min$ for the four simulations Cases LeN1 (upper left), Case LeW2 (upper right), Case HeN3 (lower left), and Case HeW4 (lower right).  
Pressures are according to the color-bars in units of $\erg \cm^{-3}$ (note different scaling of the color bars). 
}
  \label{fig:PressureVel}
    \end{figure*}

In Fig. \ref{fig:PressureVel} we present the pressure maps with stream maps (the arrows show only the direction of flow, not the speed). The two upper panels are for the low energy simulations Case LeN1 (that we analysed in section \ref{subsec:LeN1}) and Case LeW2, and the lower panels are for the high energy simulations Case HeN3 and Case HeW4. The two left panels are for narrow jets of $\alpha_{\rm j}=20^{\circ}$, and the right two panels are for wide jets of $\alpha_{\rm j}=50^{\circ}$. All panels are at $t=150 \min$ post-explosion and after 90 minutes of jet activity.
The blue zones inside the bubbles are the freely expanding pre-shock jets. 

As expected, higher energy jets (lower two panels) inflate larger bubbles. The jets' energy is 3.7 times larger in the high-energy cases, and at $t=150 \min$ their size along the z-direction is about 1.4 larger than that of the  low-energy cases. 
 
In the cases with wide jets, two right panels, the much larger jets' cross section implies much lower jets' momentum flux, and the jets' termination shocks are at small distances from the origin. A case with a wider jet forms a much smaller bubble with a much higher pressure (as energy is conserved here) with respect to the narrow jets cases.    

The flow structure is similar to that in figure \ref{fig:LeN1_vel_Evolution}. The post-shock jets' gas has a complicated flow structure with a back flow and vortexes. The post-shock ejecta gas that resides closer to the bubble boundary has an ordered outward flow.  

\subsection{Case ExG5 with an extended grid}
\label{subsec:ExG5}

\begin{figure} 
\centering
\begin{tabular}{cc}
\includegraphics[trim=0.0cm 6.0cm 0cm 5.2cm ,clip, scale=0.39]{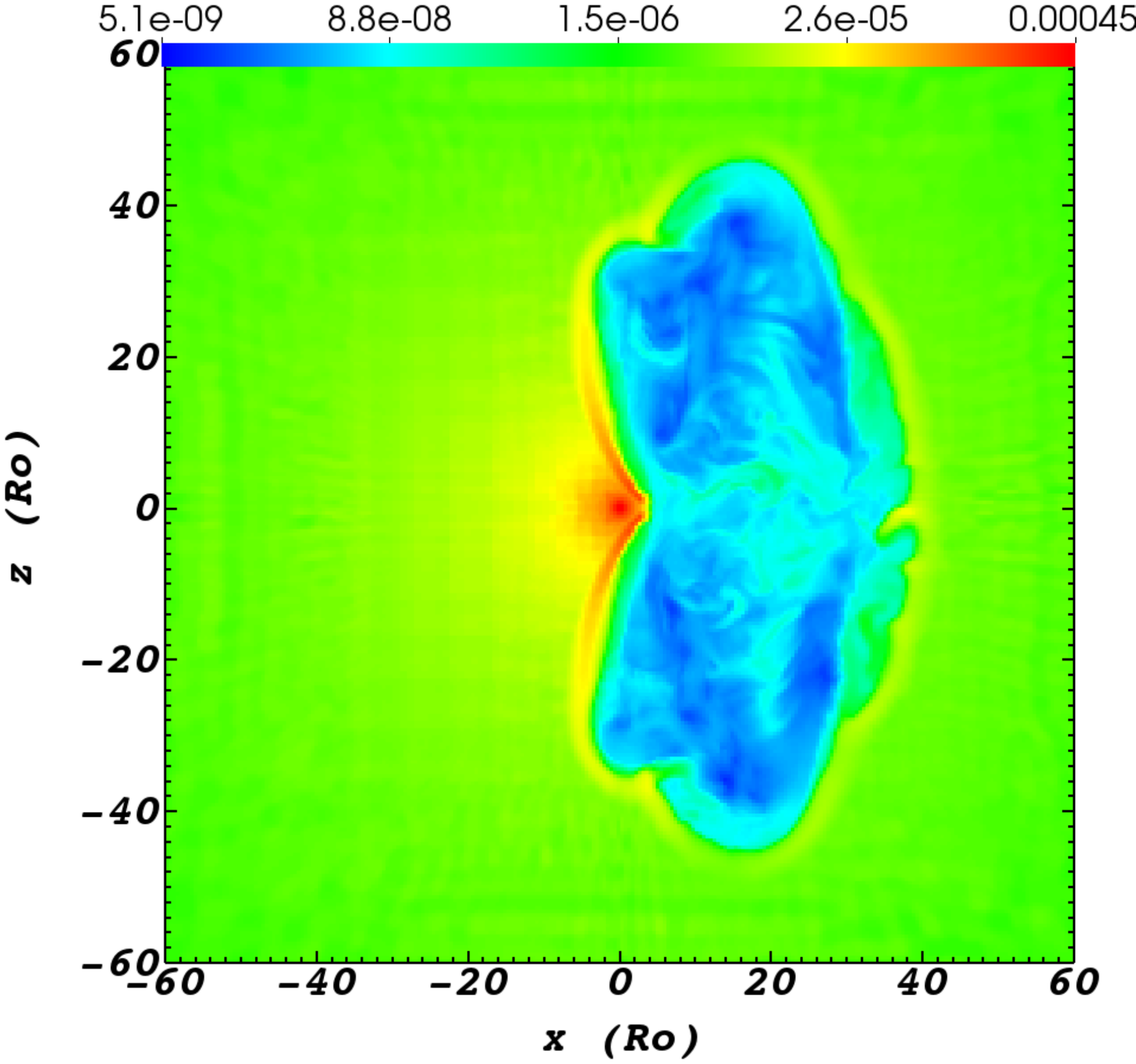} \\
\includegraphics[trim=0.0cm 6.0cm 0cm 5.2cm ,clip, scale=0.39]{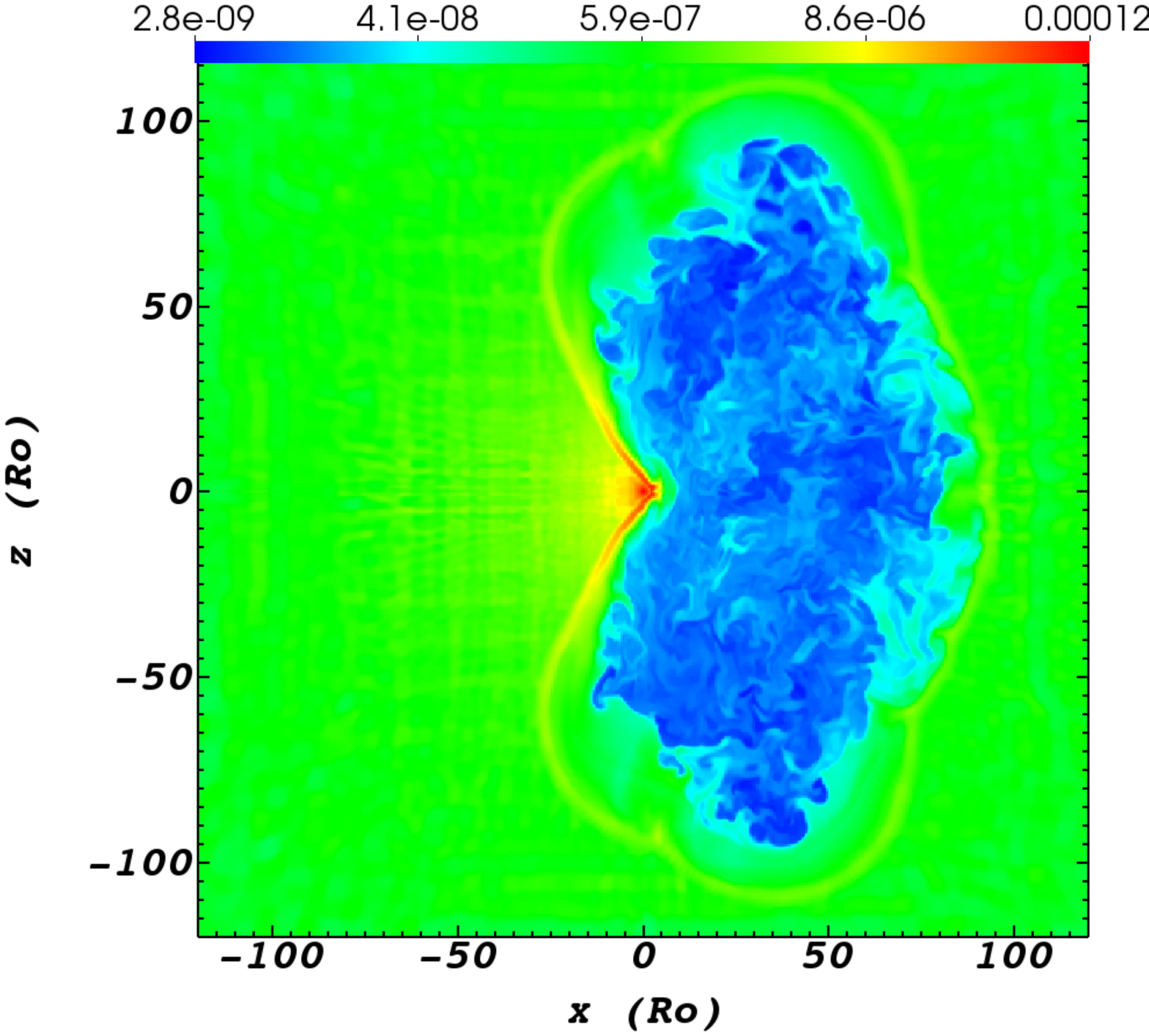}\\
\includegraphics[trim=0.0cm 6.0cm 0cm 5.2cm ,clip, scale=0.39]{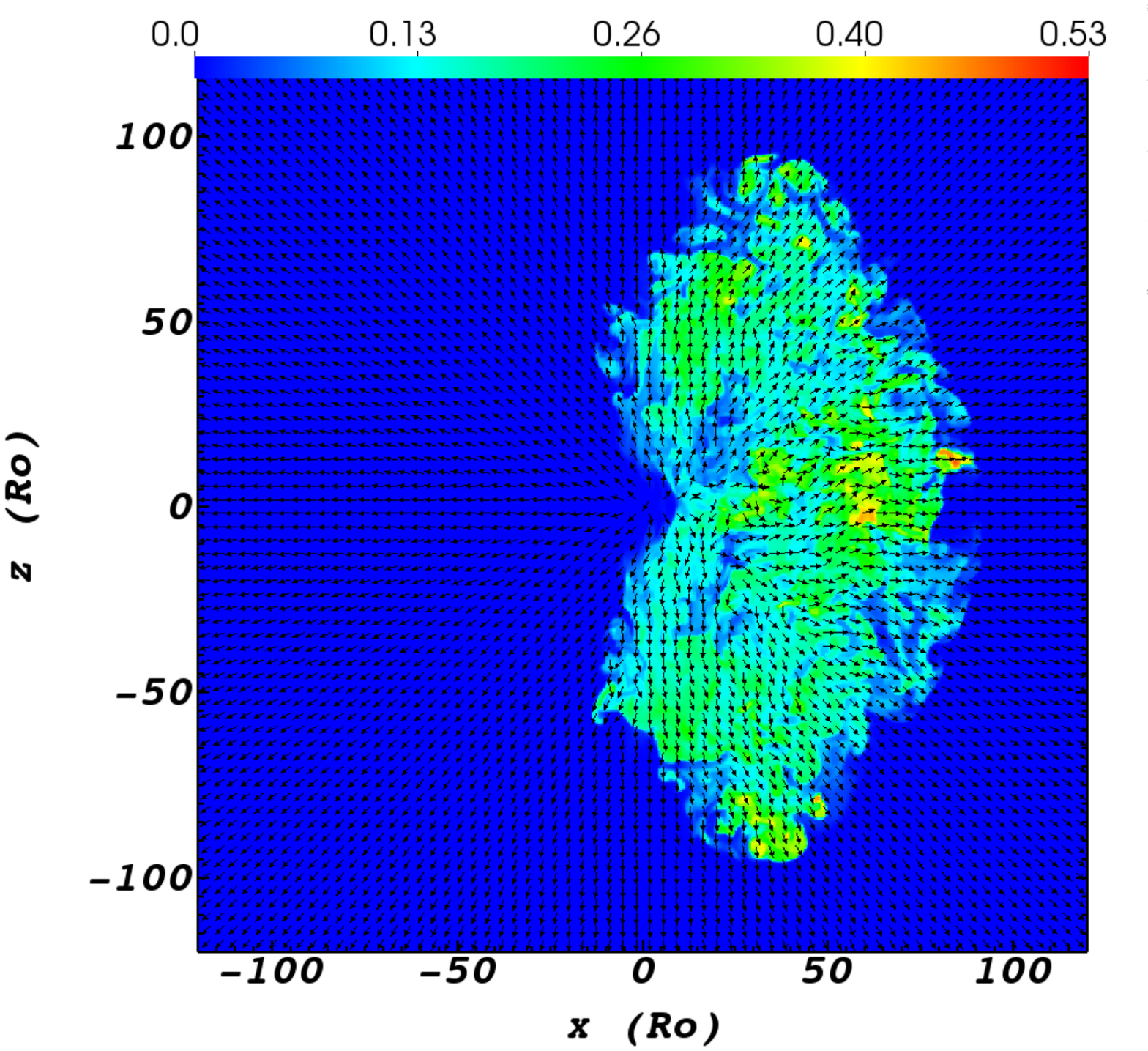}
\end{tabular}
\caption{Results of Case ExG5 with an extended numerical grid. Top: Density map in the meridional plane at $t=192 \min$. Middle: Density map in the meridional plane at $t=6.1 \h=366 \min$. Bottom: jet-tracer map at $t=6.1 \h$ with velocity vectors that depict the flow direction (but not its magnitude). 
The density scale is according to the color bar in units of $\g \cm^{-3}$. Note the increasing size of the panels with time and the different density scales of the color-bars. 
}
  \label{fig:ExG5}
    \end{figure}

Numerical resource limitations force us to reduce the numerical resolution to be able to follow the bubble for a longer time. We conduct one simulation, Case ExG5, with larger grid cells and a larger grid. We present the density maps in the meridional plane at $t=192 \min$ and $t=366 \min =6.1 \h$, and the jet-tracer map at $t=6.1 \h$ in Fig. \ref{fig:ExG5}. The physical input parameters of this simulation are as of Case LeN1. However, the larger cell sizes imply that practically the jets are somewhat wider. As we find in section \ref{subsec:OtherCases} wider jets inflate smaller bubbles at a given time. We indeed find that at $t=192 \min$ the bubble of Case  ExG5 (upper panel of Fig. \ref{fig:ExG5}) is somewhat smaller than that of Case LeN1 at the same time (lower panel of Fig. \ref{fig:LeN1_dens_Evolution}). Quantitatively, the volume of the bubble of Case ExG5 is $\simeq 0.8$ times the volume of the bubble of Case LeN1 at $t=192 \min$, or an average size ratio of $\bar a {\rm (ExG5)}=0.93 \bar a {\rm (LeN1)}$. 

We continue until the bubble reaches the edge of the numerical grid at $t= 6.1 \h$. From the lower panel of Fig. \ref{fig:ExG5} we see that the gas that originated in the jets mixed well with the ejecta mass that the bubble interacted with. 
  
The most distant parts of the bubble reach ejecta with velocities of $v_{\rm ej,i} \simeq 3500 \km \s^{-1}$. 
For the density profile that we use here (equation \ref{eq:density_profile}) the fractional mass of the ejecta with velocity of $v<v_{\rm ej,i}$ is $M(v<v_{\rm ej,i}) = (7/9)(v_{\rm ej,i}/v_{\rm br})^2 \simeq 0.2$, where in our simulations $v_{\rm br}$ is by equation (\ref{eq:vbr}). Since the bubble is onto one side of the center of explosion, the bubble in our simulations interacts and mixes with an ejecta mass of $M_{\rm ej,b} \simeq 0.1 M_{\rm ej}$. 

\section{Discussion and Summary}
\label{sec:Summary}

We simulated the interaction of jets that a NS companion to a CCSN launches from one to three hours after explosion as it accretes mass from the ejecta through an accretion disk \citep{Soker2020NSjets}. We followed the inflation of a bubble by the jets and examined the bubble morphology, size, fine detailed flow structures inside the bubble, and mixing of post-shock jets and ejecta gas. 
We placed the NS at a distance of $r=5R_\odot$, and therefore this type of flow is relevant for both SNe Ib and SNe Ic. 
We took the ejecta mass to be $M_{\rm ej}=3.5 M_\odot$, and for that our quantitative study is not relevant to ultra-stripped SNe Ic that have a much lower ejecta mass, $M_{\rm ej} \approx 0.05-0.2 M_\odot$ \citep{Taurisetal2013, Taurisetal2015, Hijikawaetal2019}. Qualitatively, though, our results are applicable even to ultra-stripped SNe Ic \citep{Soker2020NSjets}. 
  
We list our main findings as follows.
  
(1) The two bubbles that the jets inflate, one bubble by each jet, merge at a very early time to form one bubble. The ejecta pushes this bubble to the side (Fig. \ref{fig:LeN1_dens_Evolution}). Narrower jets that have a large momentum flux inflate larger bubbles that extend more along the jets' axis direction. Wider jets inflate more spherical bubbles that are smaller for the same jets' power (Fig. \ref{fig:PressureVel}).

(2) The post-shock jets' gas develops large scale meridional flows, one on each side of the equatorial plane, and small scale vortexes
(Figs. \ref{fig:LeN1_vel_Evolution} and \ref{fig:PressureVel}).
Even at the very last time, about 3 hours after we turned the jets off, we can still see a non-radial flow inside the bubble (lower panel of Fig. \ref{fig:ExG5}). 

(3) Rayleigh–Taylor instabilities develop on the contact discontinuity between the post-shock ejecta gas (the dense shell) and the post-shock jets' gas (Figs. \ref{fig:LeN1_dens_Evolution} and \ref{fig:LeN1_densXY_Evolution}). The strong flow within the bubble washes out the instability tongues (upper to lower panels in Fig. \ref{fig:LeN1_dens_Evolution}).  

(4) The instabilities and the flow inside the bubble mix postshock ejecta gas with postshock jets' gas, as we can see in the jet-tracer maps, upper panel of Fig. \ref{fig:LeN1_3D_temp_tracer} and lower panel of Fig. \ref{fig:ExG5}. 

(5) There is one case that we simulated to a late time of $t=6.1 \h$ post-explosion, Case ExG5. The most distant parts of the bubble reach ejecta with velocities of $v_{\rm ej,i} \simeq 3500 \km \s^{-1}$ (Fig. \ref{fig:ExG5}). We also present in Fig. \ref{fig:ExG5} the mixing of this material with the ejecta. By $t = 6.1 \h$ the bubble in our simulation interacts and mixes with an ejecta mass of $M_{\rm ej,b} \simeq 0.1 M_{\rm ej} \simeq 0.35 M_\odot$. 
We note that Case ExG5 is for the lower energy that we use here. As well, the jets might be active even after $t=3 \h$, although they will be weaker due to lower accretion rates. Therefore, we expect that in some cases the effects we discussed above will be stronger even. Namely, a larger bubble with a larger ejecta mass that it influences. 

Let us discuss some aspects of these results. 
\cite{Soker2020NSjets} mentions that the jets might contribute to radiation from the CCSN. 
The progenitor is a small star $R_1 \simeq 1R_\odot$. In the first few hours the ejecta expands to distances of $r_{\rm ej} \gg R_1$, and therefore suffer a large adiabatic cooling. If about half of the energy at explosion (at shock-break out from the progenitor) is thermal energy, then for an adiabatic index of $\gamma=4/3$ by the time the ejecta is at $50 R_1$ the thermal energy is only $1 \%$ of the explosion energy. The bubble suffers much less adiabatic cooling (e.g., \citealt{KaplanSoker2020} for a different setting of jets-ejecta interaction). The consequence is that although the jets' energy is $\approx 1 \%$ of the ejecta energy, the jets can add substantially to the light curve. This is particularly true for an observer on the side of the bubble, because more photons from the bubble will diffuse and escape to that direction.
Radiative transfer calculations, that we plan for the future, will have to include the calculations of opacity and of the non-spherical geometry of the problem. Our results should motivate such calculations.   
   
CCSNe lead to the nucleosynthesis of new isotopes. The distribution of the isotopes with velocity is not uniform (e.g., \citealt{Wongwathanaratetal2017}). The  instabilities and vortexes of the bubble mix these isotopes inside the bubble. 
However, the overall effect is not large, as the explosion mechnasim itself mixes these isotopes into different velocity ranges (e.g., \citealt{Wongwathanaratetal2017}). 

\cite{Soker2020NSjets} speculates that nucleosynthesis of neutron-rich material might occur in  the jets as they leave the NS companion, similar to the common envelope jets supernova (CEJS) r-process scenario (e.g.,  \citealt{GrichenerSoker2019}). We find here that if a small amount of r-process elements form, they mix with an ejecta mass of about $M_{\rm ej,b} \simeq 0.1 M_{\rm ej}$.

Our study adds to the increasing body of studies on the formation of close NS binary systems. Our study is relevant also to black hole binary systems that are formed via CCSN. Future studies that will include radiative transfer will be able to better compare to observations of CCSNe that might have extra energy in their light curves that in turn might result from a close NS companion.

\section*{Acknowledgments}
{{{{ We thank an anonymous referee for helpful comments. }}}} This research was supported by a grant from the Israel Science Foundation and a grant from the Asher Space Research Fund at the Technion.

\textbf{Data availability.}
Our setting of the FLASH code are on Zenodo:  https://zenodo.org/record/3971768\#.XylC7i2cZ0s
More information will be shared on request to the corresponding author.  

\end{document}